\documentstyle[epsf,a4,12pt]{article} 

\def\tou#1{{\lower1.2ex\hbox{$\longrightarrow$}\atop
        {\lower-.7ex\hbox{$\scriptscriptstyle #1 $}}}}
\def\lsim{{\lower1.2ex\hbox{$<$}\atop
        {\lower-.7ex\hbox{$\sim$}}}}
\def\gsim{{\lower1.2ex\hbox{$>$}\atop
        {\lower-.7ex\hbox{$\sim$}}}}

\def\be{\begin{equation}}
\def\ee{\end{equation}}

\begin{document}

\begin{titlepage}
\rightline {Si-96-13 \  \  \  \   }

\vspace*{2.truecm}

\centerline{\Large \bf  Generalized Dynamic Scaling}
\vskip 0.6truecm
\centerline{\Large \bf 
        for Critical Magnetic Systems }
\vskip 0.6truecm

\vskip 2.0truecm
\centerline{\bf B. Zheng}
\vskip 0.2truecm

\vskip 0.2truecm
\centerline{Universit\"at -- GH Siegen, D -- 57068 Siegen, Germany}

\vskip 2.5truecm

\abstract{ The short-time behaviour of
the critical dynamics for magnetic systems
  is investigated with Monte Carlo methods.
Without losing the generality, we consider the relaxation process
for the two dimensional Ising and Potts model
starting from an initial state 
with very high temperature and arbitrary magnetization.
We confirm the generalized scaling form and
 observe that the critical characteristic
functions of the initial magnetization for the Ising and the Potts model
are quite different.
}


\end{titlepage}

\section{Introduction}

The critical phenomena for magnetic systems in the equilibrium
have long been studied. Due to the large correlation length near
the critical point,
the large scale behaviour of the systems does not depend on the
microscopic details.
An universal scaling form characterized by
a number of critical exponents describes the scale
invariance in critical systems.
 For magnetic systems in {\it a non-equilibrium state},
however, it is much less understood. Indeed, for dynamic systems
in {\it the long time regime} where the equilibrium is almost reached,
one finds also a dynamic scaling form. 
Due to critical slowing down, however, numerical study
of critical dynamics is very difficult. Even for the simplest 
system as the two-dimensional Ising model,
the dynamic exponent $z$ has not rigorously been determined. 
Even though much effort has been put to develop fast algorithms
to overcome critical slowing down, the dynamic scaling has not yet
been deeply investigated.

Recently a dramatic step forward
 in the study of critical dynamics has been made.
 For a relaxation process starting from an initial state
with {\it very high temperature and small magnetization},
it is argued that universal scaling behaviour already emerges
in {\it the macroscopic early time}
\cite {jan89,jan92,oer93,oer94}.
 Important is that another dynamic
exponent should be introduced to describe the dependence of the scaling
behaviour on the initial magnetization.
It has also been discussed that under certain conditions
the effect of the initial magnetization
 remains even in the long-time regime
of the critical relaxation \cite {die93,rit96}. 
Numerical simulations support these predictions
\cite {hus89,hum91,li94,sch95,men94,oka97a,cze96,oka97,li97}.
In the equilibrium state, introducing the lattice
size $L$ as a scaling variable 
in the finite size scaling induces essential progress
in the study of critical phenomana, especially in the numerical simulation.
In dynamic systems, besides the lattice size $L$ we now
 have the evolution time $t$ as another scaling
variable.
Different from the case in the long time regime, 
it is demonstrated that in short-time dynamics the scaling variable $t$
plays a very important role in understanding the scaling
behaviour, especially in numerical simulations.
The critical exponents can be obtained from the power law behaviour 
 of the observables in the early time $t$
\cite {sch95,oka97a,sch96,sch97a} or from the 
time-dependent finite size scaling 
\cite {li95,li96}.
 Numerical results indicate that the study of
short-time dynamics not only enlarges the fundamental knowledge
on critical phenomena but also
provides new ways to determine all the
static as well as the dynamic exponents.

For the understanding of the universal behaviour in short-time dynamics,
we should keep in mind that there exist two completely different
time scales, {\it the macroscopic time scale
and the microscopic time scale}. Universal behaviour emerges only
after a time period which is long enough in the microscopic sense.
This time period is called the microscopic time scale $t_{mic}$.
We expect that $t_{mic}$ would be very small compared with
the macroscopic time scale
which is typically $t_{mac} \sim L^z$ or $t_{mac} \sim \tau ^ {-\nu z}$.
Here $z$ is the dynamic exponent and $\tau$ is the 
reduced tepemrature.
However, how big $t_{mic}$ is depends on the microscopic
details. If $t_{mic}$ would be so big as comparable
with the macroscopic time scale $t_{mac}$,
 what one claims for the universal behaviour
in the short-time dynamics becomes meaningless.

In Ref. \cite {oka97a,sch97a}, the short-time behaviour of the critical
dynamics for the two dimensional Ising and Potts model has
carefully and seriously been analysed. Simulations have been performed
with both the heat-bath and the Metropolis algorithm.
Fortunately, 
it is discovered that $t_{mic} \sim 5 - 30$
Monte Carlo time steps for both algorithms.
 If one thinks a Monte Carlo time step
is a typical microscopic time unit, this result is quite reasonable.
Taking into account the possible effect of $t_{mic}$,
critical exponents $\theta$, $z$ and $\beta/\nu$
are confidently obtained for both algorithms \cite {oka97a,sch97a}.
The results are well consistent and strongly support
universality in the short-time dynamics.
Interestingly, when the dynamic system exhibits already
universal behaviour in the macroscopic short-time regime,
the spatial correlation length is still not so big,
 e.g. for the Potts model at $t=100$,
$\xi \sim 5$ \cite {sch95}.
Some more discussions on universality in short-time dynamics
may also be found in the recent preprints 
\cite {oka97,sch97,liu95,rit96a}. 

Does the critical exponent $\theta$ (or $x_0$) descirbe
all the new physics in short-time dynamic systems?
Very recently, it has been suggested that due to
the fact that the time evolution of the magnetization
is {\it not a Markovian process}, the global
persistence exponent is also an independent critical
exponent in the dynamic systems \cite {maj96a,maj96}.
Numerical simulations show that this is indeed the case
\cite {maj96a,sch97}. The power law decay of the 
global persistence probability is directly observed and the
global persistence exponent for the two dimensional
Ising and Potts model is determined to
a satisfactory accuracy \cite {sch97}.

Since the scenario described above is so attractive
and, on the other hand,
 all the results indicate that we are still far from 
a complete understanding of the critical short-time dynamics,
we would like to push forward the study of 
more general dynamic processes.
In a previous letter \cite {zhe96},
 the author has numerically investigated
the short-time behaviour of the critical relaxation for the two 
dimensional Potts model starting from an initial state
with {\it very high temperature and arbitrary magnetization}.
A generalized dynamic scaling is found to describe the crossover
behaviour for the short-time dynamics from the repulsive
fixed point $m_0=0$
to the attractive fixed point $m_0=1$.
The critical characteristic function for the initial magnetization
is numerically determined from the finite size scaling.

The present paper is an extension of the previous letter.
We aim to present a systematic description of the numerical data,
and meanwhile pay attention to the microscopic time scale $t_{mic}$.
Compared with the data reported in the previous letter,
the maximum evolution time $t$
 here is extended to $t=2000$ for the lattice size $L=144$
and  $t=400$ for the lattice size $L=72$, twice as long as
in the previous letter \cite {zhe96}.
This allows us to see how stable the results are 
with respect to the time $t$. To reduce the fluctuations,
we have also increased the statistics for the lattice size $L=144$
for the bigger initial magnetization $m_0$.
Furthermore, as an alternative example, we have carried out also 
a similar simulation for the two dimensional Ising model.
We interestingly observe that the critical characteristic functions for
the initial magnetization for these two models are quite different.
Finally some discussions are given corresponding to a recent
preprint on this topic \cite {rit96b}.

In the next section, we present scaling analyses for the short-time dynamics
based on which the numerical data will be analysed.
In Sec. 3, numerical results for the
two dimensional Potts and Ising model will be reported.
The last section contains the conclusions.

\section{Scaling analysis}

For the critical relaxation process 
of a dynamics of the Model A starting from an initial state
with {\it very high temperature} and {\it small magnetization},
Janssen, Schaub and Schmittmann \cite {jan89}
proposed 
a dynamic scaling form which is valid up to the macroscopic
short-time regime,
\begin{equation}
M^{(k)}(t,\tau,m_{0})=b^{-k\beta/\nu}
M^{(k)}(b^{-z}t,b^{1/\nu}\tau,
b^{x_{0}}m_{0})
\label{e5}
\end{equation}
where $M^{(k)}$ is the $k$-th moment of the magnetization,
$t$ is the dynamic evolution time, $\tau$ is the reduced temperature,
and the parameter $b$ represents the spatial rescaling factor.
Besides the well known static critical exponents $\beta$, $\nu$
and the dynamic exponent $z$,
a new independent exponent $x_0$ which is
the anomalous dimension
of the initial magnetization $m_0$ is introduced
to describe the dependence of the scaling behaviour on
the initial magnetizations. Based on the scaling relation it is
predicted that at the beginning of the time evolution the magnetization
 surprisingly
undergoes a {\it critical initial increase}
\begin{equation}
M(t) \sim m_0 \, t^\theta
\label{e10}
\end{equation}
where the exponent $\theta$ is related to the exponent $x_0$ by
$\theta=(x_0-\beta/\nu)/z$.

The scaling form in Eq. 
(\ref {e5})
is valid only under the conditions that the initial state
is at very high temperature with {\it small} initial magnetization.
For a critical relaxation from
an initial state with very high temperature and
{\it arbitrary magnetization}, is there universal scaling behaviour
in the short-time regime?

In Fig. \ref {f1}, time-dependent magnetization profiles 
with different initial magnetizations $m_0$ for the critical
Potts model are displayed. The lattice size is $L=72$.
For smaller $m_0$, the magnetization continously increases.
The time scale for this incease is $t_0 \sim m_0^{-z/x_0}$.
For bigger $m_0$, the magnetization rises up rapidly and then
decreases. After a macroscopic long time $t_{mac} \sim L^z$
(out of the time range of
 Fig. \ref {f1}),
we expect that all the profiles gradually join together
and decrease exponentially. Traditionally, when we consider
the dynamic scaling in the long-time regime the effect
of the initial conditions is always neglected.
However, as we can see in Fig. \ref {f1},
in the short-time regime the effect of the intial magnetization
is very prominent, which is more or less due to the fact
the exponent $\theta$ is positive.
Therefore it is important and interesting to
study whether there exists universal behaviour in the short-time
dynamics.

A scaling form represents nothing else than a kind of scale invariance,
or self-similarity in the systems.
At the critical point the time correlation length of the 
dynamic systems is infinity, therefore
we may expect a certain scale invariance
independent of the initial conditions.
The self-similarity of the magnetization profiles
for different $m_0$ may also be observed in In Fig. \ref {f1}.
Important is that we should consider
the scaling behaviour of the initial magnetization carefully.
In the previous letter \cite {zhe96}, the author proposed a  generalized 
dynamic scaling form, e.g. 
for the $k$-th moment of the magnetization,
\begin{equation}
M^{(k)}(t,\tau,L,m_{0})=b^{-k\beta/\nu}
M^{(k)}(b^{-z}t,b^{1/\nu}\tau,b^{-1}L,
\varphi (b,m_{0}))
\label{e15}
\end{equation}
For the convenience of later discussions, finite
systems have been considered here and $L$ is the lattice size.
The scaling behaviour of the initial magnetization $m_0$
is specified by the critical characteristic function
$\varphi (b,m_{0})$, which in the limit
$m_0 \rightarrow 0$ tends to the simple
form $b^{x_{0}}m_{0}$ used in Eq.(\ref {e5}), but is in general different.
To avoid possible confusion, we have changed the notation $\chi$ 
for the critical characteristic function
in the previous letter to $\varphi$ in this paper.
This generalized scaling form describes the crossover 
behaviour between two fixed points,
the repulsive
one $m_0=0$
and the attractive one $m_0=1$.
It is also in a similar spirit
as that in the correction to the scaling,
where non-linear effects of an off-fixed point
are considered. In general cases of the crossover phenomena, however, 
$m_0$ would also enter the scaling behaviour of
the time $t$, $\tau$ and the anomalous dimension of $M^{(k)}$, or
the exponents $z$, $\beta$ and $\nu$ would depend on
$m_0$. Then the situation will be very complicated and
numerical simulations are difficult.

The simple  ansatz of the scaling form in Eq.(\ref {e15}) 
for the short-time dynamics
that the exponents $\beta$, $\nu$ and $z$
do not depend on the initial magnetization $m_0$
is based on the fact that if we believe 
there is a scaling form in the
short-time regime of the dynamic relaxation process,
this scaling form should smoothly cross over to that
in the long-time regime.
In other words, the initial magnetization should
not enter the renormalization of the critical system
in equilibrium and near equlibrium.
However, the dimension of the initial magnetization 
would, in general, depend on the value of the initial magnetization.
Assuming that we can either perturbatively or
non-perturbatively perform the renormalization calculation
of the dynamic system,
the renormalization equation for the initial magnetization $m_0$
may be written as
\begin{equation}
\frac {\partial\  \ln\ m(\lambda)} {\partial\  \ln\ \lambda} =
\eta_0(m(\lambda)).
\label{e20}
\end{equation}
Here $\eta_0(m(\lambda))$ is the anomalous dimension
of the initial magnetization and
all the irrelevant operators are already set to the fixed points.
The solution of the equation (\ref {e20}) gives 
the critical characteristic function $\varphi (b,m_{0})$.
Therefore $\varphi (b,m_{0})$ can be written in a form
\begin{equation}
\varphi (b, m_0) = F(bF^{-1}(m_0))
\label{e25}
\end{equation}
with condition $\varphi (1, m_0)=m_0$.
If we define $g=F^{-1}(m_0)$, under the renormalization 
group transformation $g$ will be scaled as $b g$.
Then we may conclude that the time scale where the effect
of $m_0$ remains is $g^{-z}$.

We should also point out that the 
dynamic scaling form in Eq.(\ref {e15})
 may straightforwardly be generalized
to a critical relaxation process starting from an initial state with
an {\it arbitrary temperature} and arbitrary magnetization.
In general the initial temperature will also enter
the dimension of the inital magnetization and vice versa.
The critical characteristic functions for the initial
magnetization and temperature are a bit more
complicated.

In order to verify the scaling form in Eq.(\ref {e15}),
as a first approach, in this paper we always set $\tau=0$. 
We perform simulations for a pair of lattice sizes $L_1$
and $L_2$. From the scaling collapse of the magnetization or
its higher moments, in principle, 
we can estimate the values of the function $\varphi (b,m_{0})$
at $b=L_2/L_1$ for different $m_0$.
This works always well in case of smaller initial magnetization.
When the initial magnetization $m_0$ is bigger,
the dynamic system leaves the repulsive fixed point $m_0=0$
and approaches to the attractive fixed point $m_0=1$.
The determination of the function $\varphi (b,m_{0})$
becomes more difficult since the dependence of the observables on 
the initial magnetization
gets weaker.
To reduce the extra fluctuations from 
$M^{(k)}(t,1)$, 
we may introduce
a magnetization difference
\begin{equation}
M_d(t,m_0)= M^{(1)}(t,1)-M^{(1)}(t,m_0)
\end{equation}
and a Binder-type cumulant
\begin{equation}
U_d(t,m_0)= \frac{M^{(2)}(t,1)-M^{(2)}(t,m_0)}{[M_d(t,m_0)]^2}
\end{equation}
to measure the function $\varphi (b,m_{0})$.

\section{Generalized dynamic scaling}

In this section, we present numerical results for the short-time dynamics
of the two-dimensional $3$-state Potts and Ising model.
In the first subsection, we systematically analyse the data for the Potts model
and demonstrate how to determine the critical characteristic function
$\varphi (b,m_0)$ from the finite size scaling.
 In the second subsection, we report the results
for the Ising model and meanwhile discuss some relevant important problems.

In this section, all the simulations and discussions are restricted to
the case $\tau=0$, i.e. the dynamic system is exactly at the critical point.
Therefore the exponent $\nu$ will not enter the calculation.

\subsection{The Potts model}

The Hamiltonian for the $q$-state Potts model is given by
\begin{equation}
H=K  \sum_{<ij>}  \delta_{\sigma_i,\sigma_j},\qquad \sigma_i=1,...,q
\label{hami}
\end{equation}
where $<ij>$ represents nearest neighbours. In our notation
the inverse temperature is already absorbed into the coupling $K$.
It is known that the critical
points locate at $K_c=log(1+\sqrt{q})$. In this paper
the $3$-state case  will be investigated.
The exact value of the static exponents
$\beta/\nu=2/15$ and the dynamic exponent $z=2.196(8)$
obtained
from the power law decay of the auto-correlations
\cite {sch95,oka97a}
will be taken as input. In the numerical simulations,
we measure the $k$th moment defined as 
\begin{equation}
M^{(k)}(t,m_0)= <\lbrack \frac{3}{2 N}\,\sum_i
      (\delta_{\sigma_i(t),1}-\frac{1}{3}) \rbrack ^{k}>, 
\end{equation}
where the average is taken over independent
initial configurations and the random forces.
The initial magnetization $m_0$ has sharply been prepared.

In order to determine the critical characteristic function
$\varphi (b,m_{0})$,  in the previous letter
we have performed the simulation
for different lattice sizes,
$L=36,72$ and $144$ with samples of different
initial configurations $80\ 000, 40\ 000$ and $8\ 000$
respectively. For the initial magnetization $m_0$,
we take the value $m_0=0.14,\ 0.22,\ 0.30,\ 0.40,\ 0.50,\
0.60,\ 0.70,\ 0.80$ and $1.00$ for the lattice size $L=144$. 
For $m_0 \ge 0.50$, the data are more fluctuating and therefore
 in this paper we increase the statistics 
for the lattice size $L=144$ to $16\ 000$ and 
for the lattice size $L=72$ to $80\ 000$.
We updated the dynamic system 
to the maximum evolution time $t=2000$ for the lattice size $L=144$,
$t=500$ for the lattice size $L=72$ and $t=100$ 
for the lattice size $L=36$.
From the scaling collapse of an observable for two different 
lattice sizes $L_1$ and $L_2$ with suitable initial magnetizations, 
we can estimate the values of the function $\varphi (b,m_{0})$
at $b=L_2/L_1$ for different $m_0$.
In principle, any observables which essentially depend
on the initial magnetization may be used for this purpose.
In this paper, the statistical errors are estimated by
dividing the samples with different initial configurations
into four groups.

In Fig. \ref {f2a}, the scaling plot for the magnetization
is displayed for 
 the lattice $L_1=72$ and the lattice $L_2=144$.
The initial magnetization for the lattice $L_2=144$
is $m_{02}=0.14$. 
If one can find an initial magnetization $m_{01}$ for lattice
$L_1=72$ such that the time evolution curves of the magnetizations
for both lattice sizes collapse, 
the scaling is valid and from
the scaling form in Eq.(\ref {e15}) one gets
$\varphi (2,m_{02})=m_{01}$.
In Fig. \ref {f2a}, 
 stars represent the time evolution of the magnetization
of the lattice $L_2=144$.
Crosses are the same data but rescaled in time and
multiplied by an overall factor $2^{\beta/\nu}$.
For the lattice $L_1=72$, 
we have performed the simulation 
with two different initial magnetizations $m_0=0.175$
and $m_0=0.185$, for which the magnetizations
have been plotted by the lower and upper solid
 lines. By linear extrapolation, we obtain
the time evolution of the magnetization with initial values
between these two values.
The solid line laying on the crosses in Fig. \ref {f2a}
 is the time evolution of
the magnetization for $L=72$
with an initial magnetization $m_{01}=0.1802(6)$ which
has the best fit to the crosses. 
Therefore, $\varphi (2,0.14)=0.1802(6)$.
Compared with the scaling plot for the magnetization difference
$M_d(t,m_0)$ presented in Fig.~1 in the previous letter,
the evolution time has been extended to $t=400$
 in the time scale of the lattice size $L=72$. 
The quality of the fit in Fig. \ref {f2a} is
as good as that in Fig.~1 in the previous letter.
The resulting value for $\varphi (2,0.14)=0.1802(6)$ is also very well
consistent with $\varphi (2,0.14)=0.1800(4)$
in the previous letter \cite {zhe96}.

However, when the initial magnetization $m_0$ is getting bigger,
the determination of the function $\varphi (b,m_0)$ 
becomes more and more difficult
since the dependence of the obsevables on $m_0$ is weakening.
For example, the time dependent magnetization
$M(t,m_0)$ can be written as $M(t,m_0)=M_d(t,m_0)+M(t,1)$.
The part depending on $m_0$ is only $M_d(t,m_0)$.
For relative big $m_0$, e.g. $m_0 \ge 0.50$,
$M_d(t,m_0)$ is much smaller than $M(t,1)$.
Therefore a small relative error of $M(t,1)$ may induce
bigger errors for the function $\varphi (b,m_0)$ when it is determined 
from the scaling plot of $M(t,m_0)$.
In Fig. \ref {f2b}, such a situation is shown for 
a scaling plot for $L=144$ with $m_0=0.50$.
The quality of the scaling collapse
is not so good. 

The errors of $M(t,1)$ have many origins.
There are apparently statistical errors.
In the rescaling of $M(t,m_0)$,
 the error of
the dynamic critical exponent $z$ also enters.
More important is the effect of the microscopic
time scale $t_{mic}$. As it is discussed
in the numerical measurements of the critical
exponent $\theta$ in Ref \cite {oka97a,sch96}, 
$t_{mic}$ becomes bigger as the initial magnetization
$m_0$ increases. For $M(t,m_0=1)$, it is observed that
$t_{mic} \sim 50$ \cite {sch96}.
Finally, the systematic errors of the random numbers
might also play some role.

As pointed out in the previous letter,
to reduce the extra errors from $M(t,1)$
one can directly estimate the funcion $\varphi (b,m_0)$
from the scaling collapse of $M_d(t,m_0)$.
In Fig. \ref {f3}, a scaling plot of $M_d(t,m_0)$
for lattice size $L=144$ with $m_0=0.50$ and lattice
size $L=72$ is plotted. Now, the fitting of the data
is much better than that for $M(t,m_0)$.
The crosses lie nicely on the fitted solid line
in the whole time interval. As it will carefully 
be discussed below,
the microscopic time scale $t_{mic}$ 
shows to be quite small here.
The resulting value of $\varphi (b,m_0)$ is
$\varphi (2,0.50)=0.6586(31)$.

In any case, errors induced by the error of 
the exponent $z$ still remain for the scaling plot
of $M_d(t,m_0)$. In the above estimation of the values
for the function $\varphi (b,m_0)$, these errors have not been
taken into account. For this reason, we have performed
the scaling fitting with two different values
of $z$, $z=2.176$ and $z=2.216$. The results
for $\varphi (2,0.50)$ are $\varphi (2,0.50)=0.6571(29)$
and $\varphi (2,0.50)=0.6602(30)$ respectively.
Compared with the value obtained with $z=2.196$,
the difference is not big. Actually all the values
are covered by the statistical errors.
In contrast to this, the result for the exponent $z$
given in Ref \cite {oka97a} is
$z=2.196(8)$ and the statistical error 
is already negligible.

In order to reveal the fine struture of the short-time
behaviour of the critical dynamics,
we can proceed in a different way to measure the
critical characteristic function $\varphi (b,m_0)$.
Especially we would like to pay attention to the possible effect
of the microscopic time scale $t_{mic}$ 
discussed in Sec. 1, and see also how stable 
the scaling form in Eq.(\ref {e15}) is  with respect 
to the evolution time.
Theoretically, the scaling form in Eq.(\ref {e15})
is valid for every time $t$ after the 
microscopic time scale $t_{mic}$. In Fig. \ref {f3},
the function $\varphi (b,m_0)$ is determined by searching
for the best fit of the data in a time interval 
$[10,400]$. We may call this {\it `global' fit}.
Actually, the function $\varphi (b,m_0)$ can be estimated
from the data in any `local' time $t$.
For example, we can perform the fitting
at a time $t$ in the time interval $[t,t+15]$ and measure
the function $\varphi (b,m_0)$. 
We may call this {\it `local' fit}.

In Fig. \ref {f4a}, \ref {f4b} and \ref {f4c},
the resulting function $\varphi (2,m_0)$ from
 the magnetization difference $M_d(t,m_0)$ for the lattice
size $L=144$ and $L=72$ as a function of the time $t$
is plotted for the initial magnetization $m_0=0.14, 0.30$ and $0.60$.
The errors are estimated by dividing the samples
with different initial configurations into four groups.
Taking into account the errors,
from the figures  we see that $\varphi (2,m_0)$ as
a function of time $t$ fluctuates around a constant value
after a microscopic time scale $t_{mic} \le 10$.
This is a strong indication that the scaling form in Eq.(\ref {e15}) 
is valid.
Such a small $t_{mic}$
 is consistent with what we observe in the numerical
measurement of the critical exponent $\theta$ \cite {oka97a}.
 The final value of $\varphi (2,m_0)$ can be estimated
by taking the average over a certain time interval.
For example, for $m_0=0.14$ we obtain $\varphi (2,0.14)=0.1801(6)$
by averaging over time interval $[10,400]$.
This coincides well with that obtained from the {\it `global' fit}.
From Fig. \ref {f4a}, \ref {f4b} and \ref {f4c},
 we can also see the tendency
that the fluctuations become bigger as $m_0$ increases.
For $m_0=0.14$ the fluctuations are around one percent
but for $m_0=0.60$ those are already around five percent.

Here we should stress that the data points of the
function $\varphi (2,m_0)$ at different time $t$
presented in Fig. \ref {f4a}, \ref {f4b} and \ref {f4c}
 are not completely statistically
independent. The data points within a time distance
of $15$ are obtained from time intervals
which partly overlap each other. Even for the data points
with a time distance bigger than $15$, they are in some way 
 correlated due to the large time
correlation length. Therefore, using the data in 
Fig.\ref {f4a}, \ref {f4b} and \ref {f4c} to estimate the
statistical errors is complicated. 
For these reasons, our statistical errors are always estimated
by dividing the samples with different initial configurations
into four groups.
However, we should also keep in mind that the data points
presented in Fig. \ref {f4a}, \ref {f4b} and \ref {f4c}
are also not completely statistically dependent.
How these data points correlate in the time direction
just represents the characteristic property of the time
evolution of the magnetization. This kind of pictures
as Fig.\ref {f4a}, \ref {f4b} and \ref {f4c} have shown to
be useful in handling the possible effects of
the microscopic time scale $t_{mic}$ and of finite lattice sizes
\cite {li96,oka97a}. 
It helps us to decide in which time interval we should
perform the {\it `global' fit} to obtain the final results
for the function $\varphi (b,m_0)$.

The function $\varphi (b,m_0)$ can also be obtain from 
other observables. In Fig.\ref {f5}, a scaling plot 
is shown for the Binder-type cumulant $U_d(t,m_0)$
for the lattice size $L=144$ with $m_0=0.14$ and the lattice size
$L=72$. The result $\varphi (2,0.14)=0.1795(8)$ is as good as that 
estimated from $M_d(t,m_0)$.
To determine the function $\varphi (b,m_0)$ from the
cumulant $U_d(t,m_0)$, one does not need the
static exponent $\beta/\nu$ as an input.
This is an advantage. The fact that $\varphi (b,m_0)$
obtained from both $M_d(t,m_0)$ and $U_d(t,m_0)$ coincide
well provides strong support for the generalized
scaling form in Eq.(\ref{e15}).
However, for bigger initial magnetization, e.g. $m_0 \ge 0.50$,
 $U_d(t,m_0)$ fluctuates much more than $M_d(t,m_0)$ does.

In Table \ref {t1}, the functions $\varphi (2,m_0)$ and
$\varphi (4,m_0)$ obtained from a global fit of
the magnetization difference $M_d(t,m_0)$ are listed.
To estimate $\varphi (2,m_0)$, 
the fitting is carried out in a time interval $[10,200]$
for smaller $m_0$,
while in a time interval $[10,100]$ for $m_0 > 0.50$.
For $\varphi (4,m_0)$ the fitting is performed 
in a time interval $[10,100]$
for smaller $m_0$,
while in a time interval $[10,50]$ for $m_0 > 0.50$.
The errors are estimated by dividing the samples
with different initial configurations into four groups.
Compared with the data in the previous letter
the results for bigger $m_0$ are improved.

In order to get a more direct understanding
 of the full critical characteristic function
$\varphi (b,m_{0})$, in the previous letter we have  defined
an effective dimension $x(b,m_0)$ of the initial magnetization
$m_0$ by $\varphi (b,m_{0}) = b\ ^{x(b,m_0)}\ m_0$.
Taking the results from Table \ref {t1}, the corresponding
effective dimensions $x(b,m_0)$ are plotted in Fig.  \ref {f6}.
Circles
 are of $x(2,m_0)$
and crosses  are of $x(4,m_0)$. 
They clearly show that the function $\varphi (b,m_{0})$ is
not a constant with respect to $m_0$. Later we will come
back to this point at the end of the next subsection.

Before closing this subsection, we would like to
discuss on the technique
of {\it the sharp preparation of
the initial magnetization}
in the numerical simulations.
Taking the Potts model as an example,
 we randomly generate
the initial spin configurations such that
 each spin takes value $+1$
with probability $(1+2m_0)/3$, while
 $0$ and $-1$ with probability $(1-m_0)/3$.
If the lattice size is infinite, a {\it sharp} value $m_0$
for the initial magnetization
is automatically achieved. However, practically our lattice
size is finite. Then the initial magnetization obtained in this
way fluctuates around the value $m_0$, depending on the
lattice size $L$. This is an extra kind of finite size effects.
To reduce this effect,
we usually adopt the sharp preparation technique:
when the initial magnetization does not take the value $m_0$,
we randomly select
 a spin on the lattice and flip it if the updated
 magnetization comes nearer to $m_0$; we continue this procedure until
 the initial magnetization reaches the value $m_0$.
In the previous numerical simulations, this technique has
 been widely used and shown its efficiency.
\cite {li94,sch95,sch96,li95,li96,rit95}.
For sufficient big lattice sizes,
whether one applies the sharp preparation technique does not
make a big difference, but the data with sharply prepared
$m_0$ show in general less extra finite size effects.
How important the sharp preparation technique is depends
also on the initial magnetization $m_0$ and what kind
of observables one measures. The smaller (or nearer to one)
the initial magnetization $m_0$ is, the more important
sharp preparation technique becomes. In two very recent
preprints \cite {oka97,sch97}, more detailed discussions
on this problem have been presented. A typical example
is the measurement of the global persistence exponent
\cite {maj96a,sch97}. Without the sharp preparation
technique, it is very difficult to observe the power law decay
of the persistence probability of the (global) magnetization
and the extra finite size effect remains even 
when the lattice size $L=256$ for the two dimensional Ising model
\cite {sch97}.

For the measurement of the critical characteristic
function, we have also repeated some calculations
without the sharp preparation of the initial magnetization.
In Fig. \ref {f7}, a scaling plot for the magnetization difference
$M_d(t)$ is displayed for the lattice size 
$L=72$ and $L=144$. The initial magnetization 
for $L=144$ is $m_0=0.14$. Stars represent
the time evolution of the magnetization difference
for $L=144$, while the lower and upper solid line
are those for $L=72$ with $m_0=0.175$ and $m_0=0.185$
respectively. For comparison, the correponding
magnetizations for the lattice size $L=144$ and $L=72$
with the sharp preparation of the initial magnetization
have also been displayed by the dotted line and
the dashed lines in the figure. We see the difference 
between with and without the sharp preparation technique
is already not big.
The crosses are the same data as the stars
 but rescaled in time. The resulting
value of $\varphi (b,m_{0})$ is
$\varphi (2,0.14) = 0.1811(9)$.
Within the statistical errors,
this is consistent with $\varphi (2,0.14) = 0.1800(4)$
obtained with the sharp preparation technique.
Similarly, without the sharp preparation technique
we obtain $\varphi (2,0.40) = 0.5283(17)$. Looking at 
Table \ref {t1}, all these results show that
for the lattice as big as $L=72$ and $L=144$
the difference between with and without the sharp preparation
technique is within the statistical errors.

\subsection{The Ising model}

The Hamiltonian for the Ising model is
\begin{equation}
H=J  \sum_{<ij>}  S_i\ S_j\;,\qquad S_i=\pm 1\;,
\label{hamii}
\end{equation}
with $<ij>$ representing nearest neighbours.
In equilibrium the Ising model is exactly solvable.
The critical point locates at $J_c=\log(1+\surd 2)/2$
and the exponent $\beta/\nu=1/8$.
The $k$th moment is defined as
\begin{equation}
M(t)= \langle \lbrack\frac{1}{N}\,\langle \sum_i S_i(t) \rbrack ^{k} \rangle.
\end{equation}

In this section, we follow the same procedure as demonstrated
in the last subsection to study the short-time behaviour of
the critical dynamics  for the Ising model 
and to determine the critical characteristic function
for the initial magnetization.
It has been pointed out in the previous letter \cite {zhe96},
 for the Potts model
 in the determination of $\varphi (2,m_{0})$
the finite size effect for the lattice pair $(L_1,L_2)=(72,144)$
is already very small. This situation is similar for the Ising model.
Therefore, in this section we restrict our discussions
to the lattice pair $(L_1,L_2)=(64,128)$. 
For the determination of $\varphi (4,m_{0})$, we use the lattice pair
$(L_1,L_2)=(32,128)$.
For lattice sizes $L=36,72$ and $144$, we have performed simulations
with  $80\ 000, 32\ 000$ and $8\ 000$
samples of different initial configurations
respectively. 
The initial magnetization for the lattice $L=128$ takes the value 
$m_0=0.10,\ 0.20,\ 0.30,\ 0.40,\ 0.50,\
0.60,\ 0.70,\ 0.80$ and $1.00$.
For the lattice $L=128$ with $m_0 \ge 0.60$,
the statistics is  $16\ 000$.
The maxmum evolution time is 
$t=2500$ for the lattice size $L=128$,
$t=250$ for the lattice size $L=72$ and $t=100$
for the lattice size $L=36$.

In case of the Potts model, we believe that the dynamic
exponent $z=2.196$ used in the last subsection
is very close to the true value
\cite {sch95,oka97a,sch96}.
However, for the Ising model the value of the dynamic exponent
is still a matter of debate. Values ranging 
from $z=2.13$ to $z=2.17$ can be found in the literature
\cite {tan87,lan88,ito93,li95,li96,gra95,nig96}.
Fortunately, in the determination of the critical characteristic
function $\varphi (b,m_{0})$ the deviation about one percent
in the value of $z$ does not induce a big change in the results
for $\varphi (b,m_{0})$, at least for not too big $m_0$.
In Fig. \ref{f8}, a scaling plot for the magnetization of
the lattice size $L=128$ with $m_0=0.40$ and lattice size
$L=64$ is displayed. Crosses and circles are the same data for
the magnetization of $L=128$, but rescaled with two  
different values of the dynamic exponent $z$,
 $z=2.15$ and $z=2.17$ respectively.
We can see the difference is very small.
The resulting values of $\varphi (2,m_{0})$ are 
$\varphi (2,0.40)=0.5588(25)$ for $z=2.15$ and
$\varphi (2,0.40)=0.5567(25)$ for $z=2.17$.
On the other hand, in our numerical measurements from the short-time
dynamics the value of $z$ tends to $z=2.15$ \cite {li96,oka97a}.
Therefore, in this section we always use the value $z=2.15$
in the analyses of the data unless we specify other values.
Later we will give some more details on this point.

Here we should also mention that as in the case of
the Potts model, for the lattice pair
$(L_1,L_2)=(64,128)$ the difference between data
with and without the sharp preparation technique,
 i.e. the extra finite size effect from the initial configurations
is already not big. For example, without the sharp preparation
of the initial magnetization we get 
$\varphi (2,0.40)=0.5566(18)$ from the scaling collaps of the
magnetization. Wihtin the statistical errors, this is consistent with 
$\varphi (2,0.40)=0.5588(25)$ obtained with the 
sharp preparation technique.

In Fig. \ref{f9}, a scaling plot for the magnetization of
the lattice size $L=128$ with $m_0=0.30$ and lattice size
$L=32$ is displayed. The scaling collapse looks
also very nice even though the fluctuation is slightly
bigger than that in the scaling plot for the lattice pair
$(L_1,L_2)=(64,128)$. The resulting value of $\varphi (2,m_{0})$ 
is $\varphi (4,0.30)=0.585(2)$.
Practically, the measurement of $\varphi (4,m_0)$ is more difficult
than that of $\varphi (2,m_0)$. We have to perform simulations
for bigger lattice sizes and longer evolution time.
Much computer time is needed. Therefore, to measure $\varphi (b,m_0)$
for even bigger $b$, it is not wise to follow the way
discussed in this paper.

To see the possible effect of the microscopic time scale 
$t_{mic}$, we can analyse the data also by the 
local fit. In Fig. \ref{f10}, $\varphi (2,m_{0})$
with $m_0=0.40$ obtained from the magnetization with
lattice size $L=128$ and $L=64$ is plotted versus time $t$.
We can see clearly that $t_{mic}$ is also very small,
$t_{mic} \le 10$. Therefore, in this section we always start
 our measurements of the critical characteristic function
$\varphi (2,m_{0})$ from the time $t=10$.

In case of the Potts model, when $m_0$ is bigger 
the scaling collapse for
the magnetization $M(t,m_0)$ is not as good as 
that for the magnetization difference $M_d(t,m_0)$.
However, it is interesting that for the Ising model
both scaling plots for $M(t,m_0)$ and $M_d(t,m_0)$
give similar results, at least when one takes $z=2.15$.
 This shows that in the simulations
we do not get so much fluctuations from $M(t,1)$.
In other words, the data of $M(t,1)$ for the lattice pair
$(L_1,L_2)$  collapse very well.
In Table \ref {t2}, the function $\varphi (2,m_0)$ and
$\varphi (4,m_0)$ obtained from a global fit of
the magnetization  $M(t,m_0)$ are listed.
For $z=2.17$, the case is slightly different.
We will discuss this later.

In Fig. \ref{f6}, the final results for the effective dimension
$x (2,m_{0})$ and $x (4,m_{0})$ are
given. The data are taken from Table \ref{t2}.
Stars with solid line
 are of $x(2,m_0)$ and diamonds with dashed line are of $x(4,m_0)$.
From  Fig. \ref{f6} we observe that
the  effective dimension $x (b,m_{0})$
for the Ising model and the Potts model is quite different.
We would expect that $x (b,m_{0})$
always monotonously decreases as $m_0$ goes from $m_0=0$
to $m_0=1$. This is indeed the case for the Ising model.
However, for the Potts model $x (b,m_{0})$
first increases and after some time reaches its maximum
and then decreases.
This is consistent with what one observes in the numerical measurement
of the exponent $\theta$ \cite {oka97a}.
When $m_0$ increases from $m_0=0.0$, for the Ising model
the effective $\theta$ decreases while for the Potts model
it increases.
This phenomanon remains to be understood.
It is challenging to derive analytically the generalized
dynamic scaling form in Eq.(\ref{e15}) and the critical
characteristic function $\varphi (b,m_0)$.

Now, naturally one may ask, is the critical characteristic
function $\varphi (b,m_{0})$ really non-trivial?
In other words, it may happen that if we take another
field rather than the initial magnetization
as the scaling variable one exponent is sufficient to describe
the scaling behaviour of the initial magnetization.
{\it Theoretically}, this is possible. For example, we can take
$g=F^{-1}(m_0)$ as a scaling variable as it is discussed 
in Sec. 2. However, the problem is not so simple.
First, in general we don't know the function $F^{-1}(m_0)$.
Secondly, the physical meaning of $g$ is not clear.
Actually, $F^{-1}(m_0)$ is just the function we 
intend to measure.
Very recently, we have received a preprint \cite {rit96b}
by U. Ritschel and P. Czerner
(Univ. Essen,FRG) 
where a comment is given to the topic
addressed in this paper and the previous letter \cite {zhe96}.
It is said that if one uses the initial external magnetic field
as the scaling variable rather than the initial magnetization,
the exponent $x_0$ is sufficient to describe the dynamic scaling.
They got an indication for that from their data
of $x(b,m_0)$ for the Ising model.
However, their conjecture apparently fails in the Potts model.
For the Ising model, their data  overlap partly and are consistent
with ours within the errors. But our lattice sizes are bigger
and our errors are smaller.
Our data within errors are not consistent with
 their conjecture.
Actually,
some points of their data are also outside of
their expected values.

At high temperature,
the external magnetic field $h$ and the initial magnetization
$m_0$ have a simple relation,
for the Ising model \cite {rit96b}
\begin{equation}
m_0=\tanh(h),
\end{equation}
and for the Potts model
\begin{equation}
m_0=\frac{3}{2}(\frac{e^h}{e^h+2}-\frac{1}{3}).
\end{equation}
Supposing $h$ is used as a scaling variable 
in the scaling form in Eq.(\ref {e15}),
the effective anomalous dimension $y(b,m_0)$
of $h$ can also be extracted from the numerical data.
In the limit $m_0=0.0$, $y(b,0)=x(b,0)=x_0$. 
The exponent $x_0$, which 
has been measured 
in Ref. \cite {sch95,oka97a}, is $x_0=0.536$ and $x_0=0.298$
for both the Ising and the Potts model respectively.
The results are shown in Fig. \ref {f11}.
For the Potts model, $y(b,m_0)$ apparently is not a constant.
For the Ising model, it is not too far from a constant but
there is significant difference.
Especially, as is pointed out before,
 the results obtained from both $M_d(t,m_0)$ and $M(t,m_0)$
agree very well.
For $m_0 \ge 0.80$, the fluctuations for the Potts model
 become very big.
We can not see whether $y(2,m_0)$ continues to increase or
starts to decrease. 
For the Ising model, the data are not available.

We have discussed  that the dynamic exponent $z$ is not known so 
rigorously. Therefore we may also present the results with
other values of $z$. For example, if one takes 
$z=2.17$ for the Ising model, the resulting $y(2,m_0)$
is displayed in Fig. \ref {f12}.
Stars are of $y(2,m_0)$ measured from $M(t,m_0)$ and squares 
are that from $M_d(t,m_0)$. 
The results obtained from $M_d(t,m_0)$ coincide very well with
those measured with $z=2.15$ (the solid line).
However, for relative bigger $m_0$ the data from $M(t,m_0)$
show some deviation. We believe the result of $M_d(t,m_0)$
are nearer to the true values of $y(2,m_0)$.
In the scaling plot for $M(t,m_0)$ one may have got some errors
from $M(t,1)$. This might also indicate that $z=2.15$
is a better value for the dynamic exponent $z$ of the Ising model.
Anyhow, even the data from $M(t,m_0)$ are not compatable with
a constant.

\section{Conclusions and remarks}

We have numerically simulated the universal
short-time behaviour
of the dynamic relaxation process for the two dimensional
 Ising and the Potts model at criticality starting from an initial state
with very high temperature and {\it arbitrary} magnetization.
The critical characteristic function for the initial
magnetization, which is introduced to describe the
generalized dynamic scaling, is determined
with finite size scaling.
It is interesting that the critical characteristic functions
for the Ising and the Potts model look quite different.
When the initial magnetization varies from $m_0=0$ to $m_0=1$,
the effective dimension of the initial
magnetization for the Ising model
monotonously decreases, while that for the Potts model
first rises and then decreases.
It is challenging to derive analytically the generalized scaling
form as well as the critical characteristic
function $\varphi (b,m_{0})$. However, 
the standard perturbative renormalization group methods for the
continuum models,  as  e.g. the $\phi ^4$-theory, do not
apply here. 

The initial magnetization can be considered as a coupling of the systems.
The generalized scaling form describes the crossover scaling behaviour
of the dynamic systems between the fixed point $m_0=0$ to $m_0=1$.
This kind of phenomena can be found in many statistical systems,
for example, at the ordinary phase transition in the surface
critical phenomena. When a non-zero magnetization is applied to the
surface, a similar scenario as in the short-time dynamics appears.

{\it Acknowledgement:} The author would like to thank
   L. Sch\"ulke for helpful discussions.


\newpage

\begin{table}[h]\centering
$$
\begin{array}{|c|l|l|}
\hline
     m_0  & \chi (2,m_0) &  \chi (4,m_0) \\
\hline
    0.14  &  0.1800(04) &   0.233(1)   \\
    0.22  &  0.2866(10) &   0.378(2)  \\
    0.30  &  0.3952(11) &   0.521(3)  \\
    0.40  &  0.5299(07) &   0.691(4)  \\
    0.50  &  0.6586(31) &   0.815(8)  \\
    0.60  &  0.7678(23) &   0.907(6)  \\
    0.70  &  0.8657(29) &   0.954(3)  \\
    0.80  &  0.9272(91) &   0.987(7)  \\
\hline
\end{array}
$$
\caption{
 The characteristic function $\varphi (b,m_0)$ for the Potts model measured
from the magnetization difference $M_d(t)$.
}
\label{t1}
\end{table}

\begin{table}[h]\centering
$$
\begin{array}{|c|l|l|}
\hline
     m_0  & \varphi (2,m_0) &  \varphi (4,m_0) \\
\hline
    0.10  &  0.1439(04) &   0.206(01)   \\
    0.20  &  0.2857(14) &   0.405(03)  \\
    0.30  &  0.4254(10) &   0.585(02)  \\
    0.40  &  0.5588(25) &   0.741(07)  \\
    0.50  &  0.6800(31) &   0.847(13)  \\
    0.60  &  0.7890(38) &   0.931(13)  \\
    0.70  &  0.8783(78) &   0.964(12)  \\
\hline
\end{array}
$$
\caption{
 The characteristic function  $\varphi (b,m_0)$ for the Ising model measured
from the magnetization.
}
\label{t2}
\end{table}

\newpage

\renewcommand{\thefigure}{1}

\begin{figure}[t]\centering
\epsfysize=12cm
\epsfclipoff
\fboxsep=0pt
\setlength{\unitlength}{1cm}
\begin{picture}(13.6,12)(0,0)
\put(1.2,8.0){\makebox(0,0){$M(t)$}}
\put(12.8,1.2){\makebox(0,0){$t$}}
\put(0,0){{\epsffile{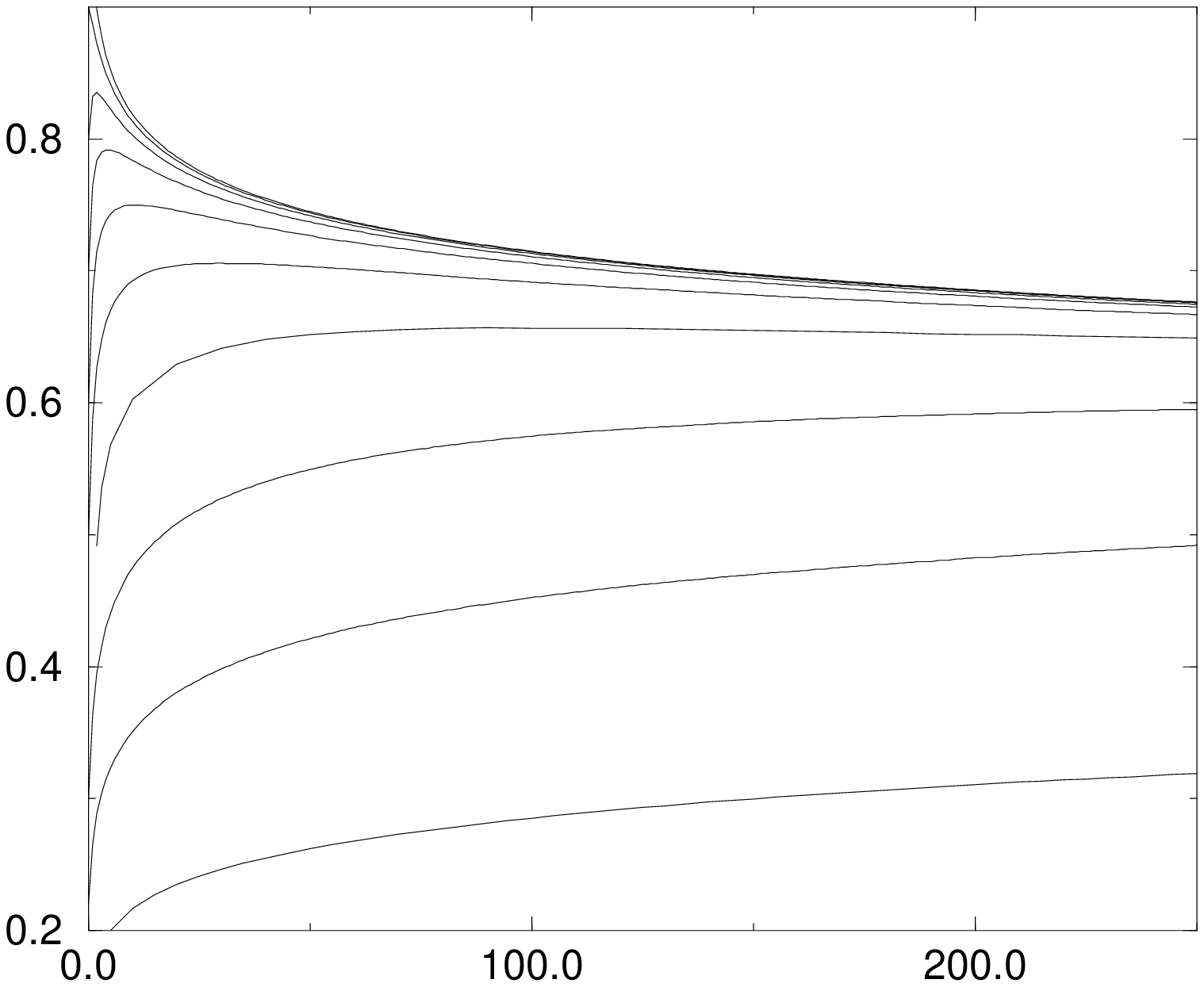}}}
\end{picture}
\caption{ Magnetization profiles for the Potts model
with $L=72$ and $m_0=0.14,\ 0.22,\ 0.30,\ 0.40,\ 0.50,\
0.60,\ 0.70,\ 0.80,\ 0.90$ and $1.00$.
}
\label{f1}
\end{figure}

\renewcommand{\thefigure}{2a}

\begin{figure}[t]\centering
\epsfysize=12cm
\epsfclipoff
\fboxsep=0pt
\setlength{\unitlength}{1cm}
\begin{picture}(13.6,12)(0,0)
\put(1.2,8.0){\makebox(0,0){$M(t)$}}
\put(11.8,1.2){\makebox(0,0){$t$}}
\put(8.4,5.5){\makebox(0,0){\footnotesize$L=144, m_0=0.14$}}
\put(10.,8.){\makebox(0,0){\footnotesize$L=72, m_0=0.175$}}
\put(7.2,10.){\makebox(0,0){\footnotesize$L=72, m_0=0.185$}}
\put(0,0){{\epsffile{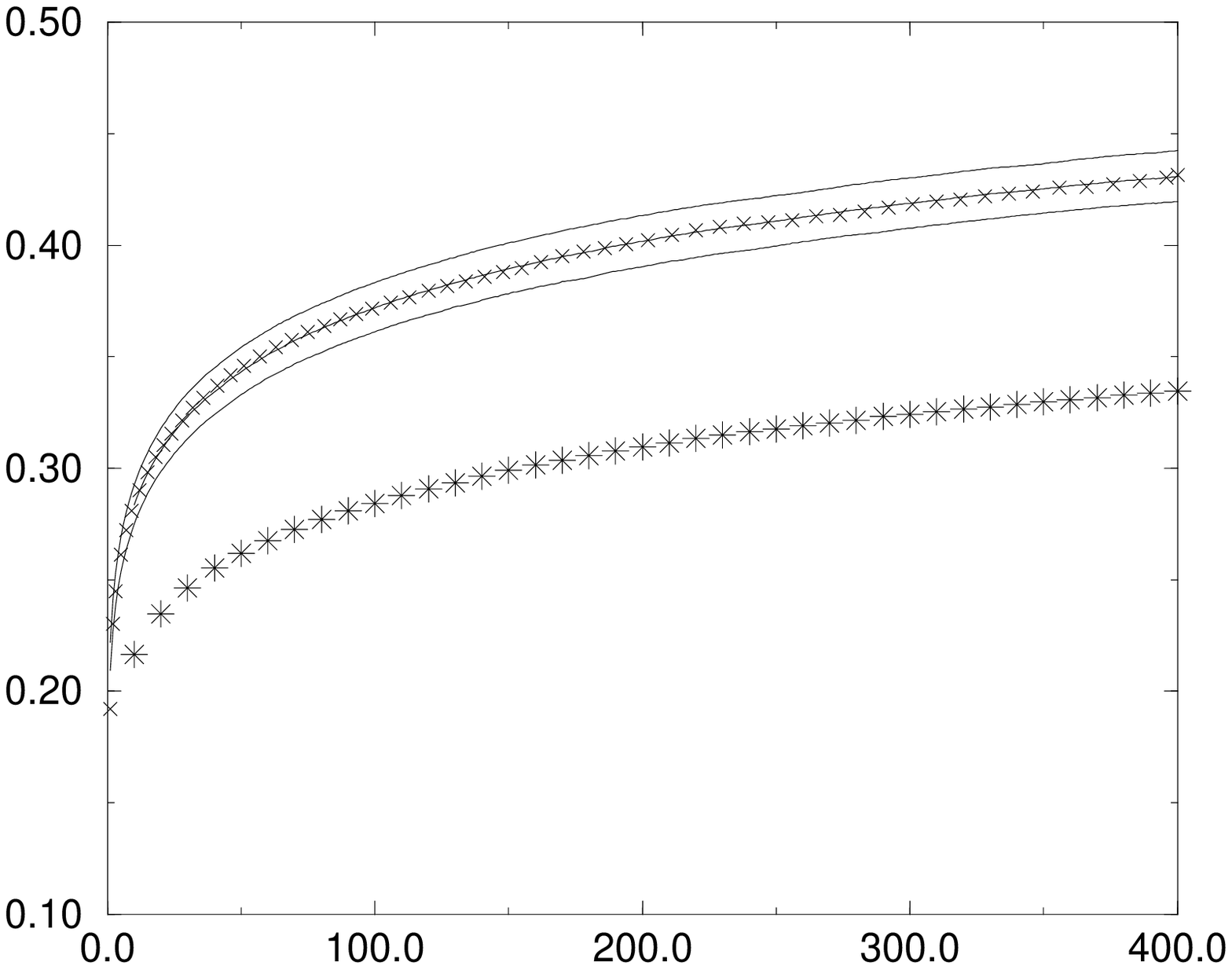}}}
\end{picture}
\caption{ The scaling plot for the magnetization with $(L_1,L_2)=(72,144)$
for the Potts model.
For the lattice size $L_2=144$, the initial magnetization $m_0=0.14$.}
\label{f2a}
\end{figure}

\renewcommand{\thefigure}{2b}

\begin{figure}[t]\centering
\epsfysize=12cm
\epsfclipoff
\fboxsep=0pt
\setlength{\unitlength}{1cm}
\begin{picture}(13.6,12)(0,0)
\put(1.2,8.0){\makebox(0,0){$M(t)$}}
\put(11.8,1.2){\makebox(0,0){$t$}}
\put(5.,2.6){\makebox(0,0){\footnotesize$L=144, m_0=0.50$}}
\put(4.,5.5){\makebox(0,0){\footnotesize$L=72, m_0=0.60$}}
\put(8.2,7.){\makebox(0,0){\footnotesize$L=72, m_0=0.70$}}
\put(0,0){{\epsffile{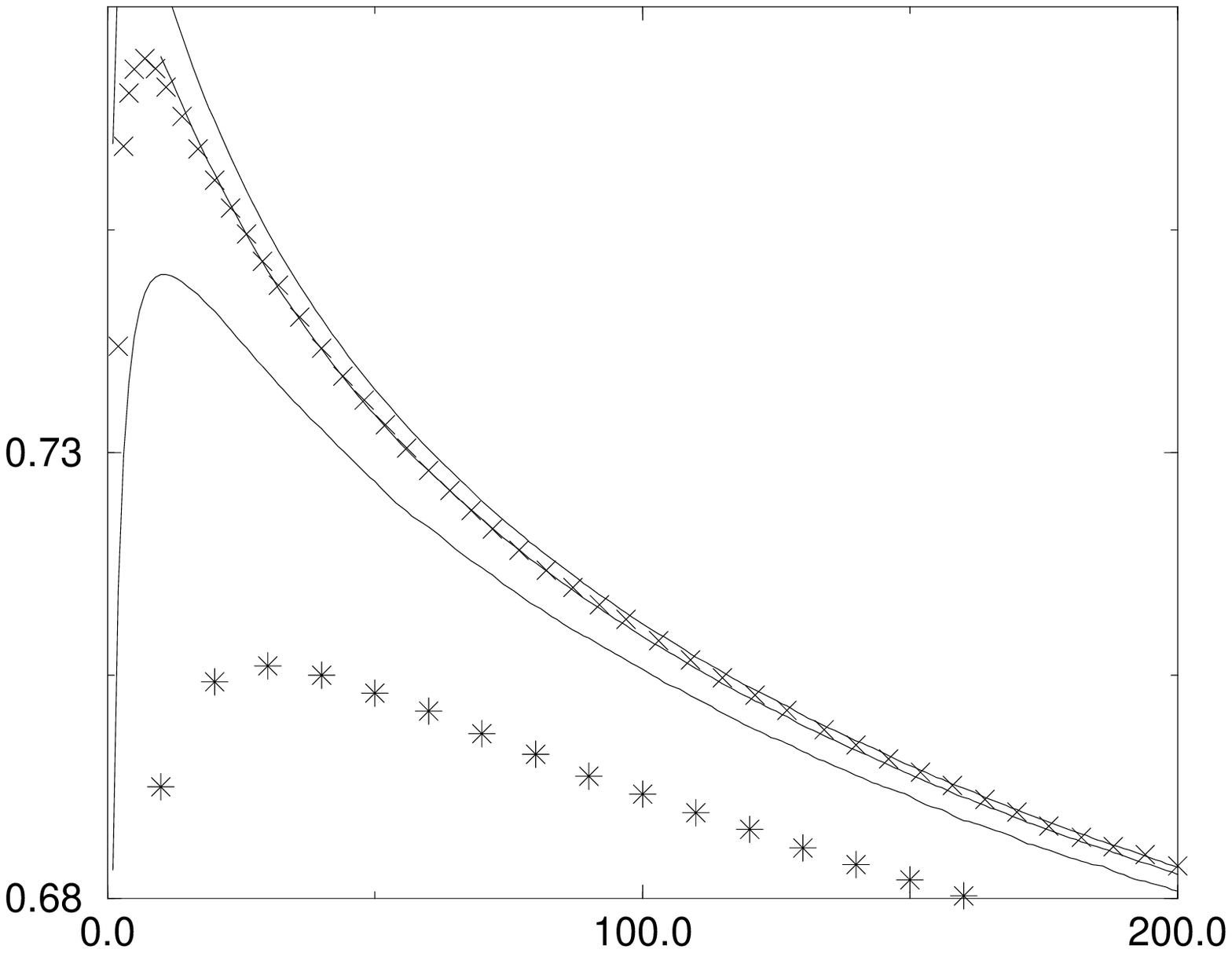}}}
\end{picture}
\caption{ The scaling plot for the magnetization with $(L_1,L_2)=(72,144)$
for the Potts model.
For the lattice size $L_2=144$, the initial magnetization $m_0=0.50$.}
\label{f2b}
\end{figure}

\renewcommand{\thefigure}{3}

\begin{figure}[t]\centering
\epsfysize=12cm
\epsfclipoff
\fboxsep=0pt
\setlength{\unitlength}{1cm}
\begin{picture}(13.6,12)(0,0)
\put(1.2,8.0){\makebox(0,0){$M_d(t)$}}
\put(11.8,1.2){\makebox(0,0){$t$}}
\put(7.,7.6){\makebox(0,0){\footnotesize$L=144, m_0=0.50$}}
\put(6.7,3.7){\makebox(0,0){\footnotesize$L=72, m_0=0.60$}}
\put(4.,2.3){\makebox(0,0){\footnotesize$L=72, m_0=0.70$}}
\put(0,0){{\epsffile{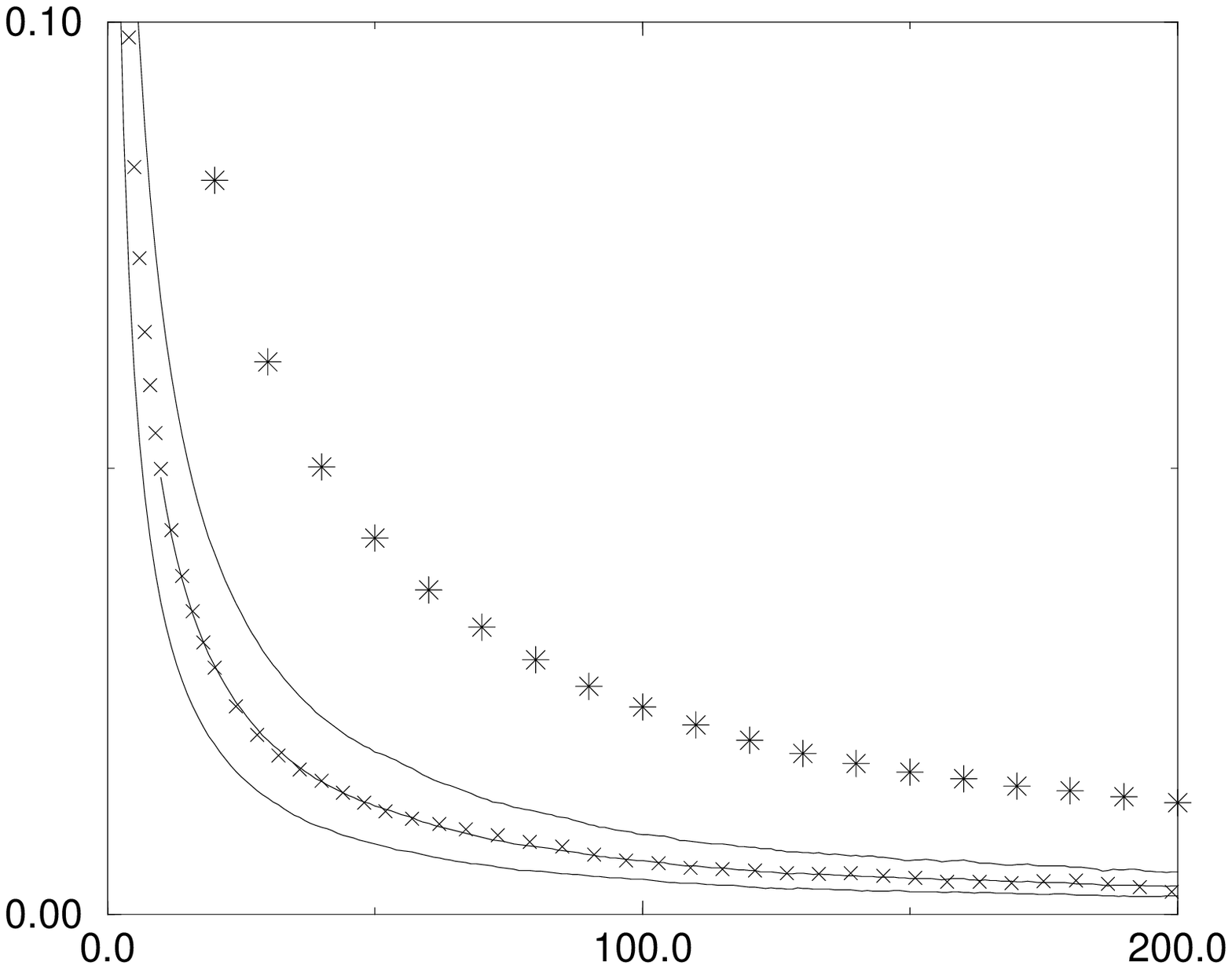}}}
\end{picture}
\caption{ The scaling plot for the magnetization 
difference $M_d(t)$ with $(L_1,L_2)=(72,144)$ for the Potts model.
For the lattice size $L_2=144$, the initial magnetization $m_0=0.50$.}
\label{f3}
\end{figure}

\renewcommand{\thefigure}{4a}

\begin{figure}[t]\centering
\epsfysize=12cm
\epsfclipoff
\fboxsep=0pt
\setlength{\unitlength}{1cm}
\begin{picture}(13.6,12)(0,0)
\put(1.2,9.0){\makebox(0,0){$\varphi(2,m_0)$}}
\put(11.8,1.2){\makebox(0,0){$t$}}
\put(6.7,5.2){\makebox(0,0){\footnotesize$m_0=0.14$}}
\put(0,0){{\epsffile{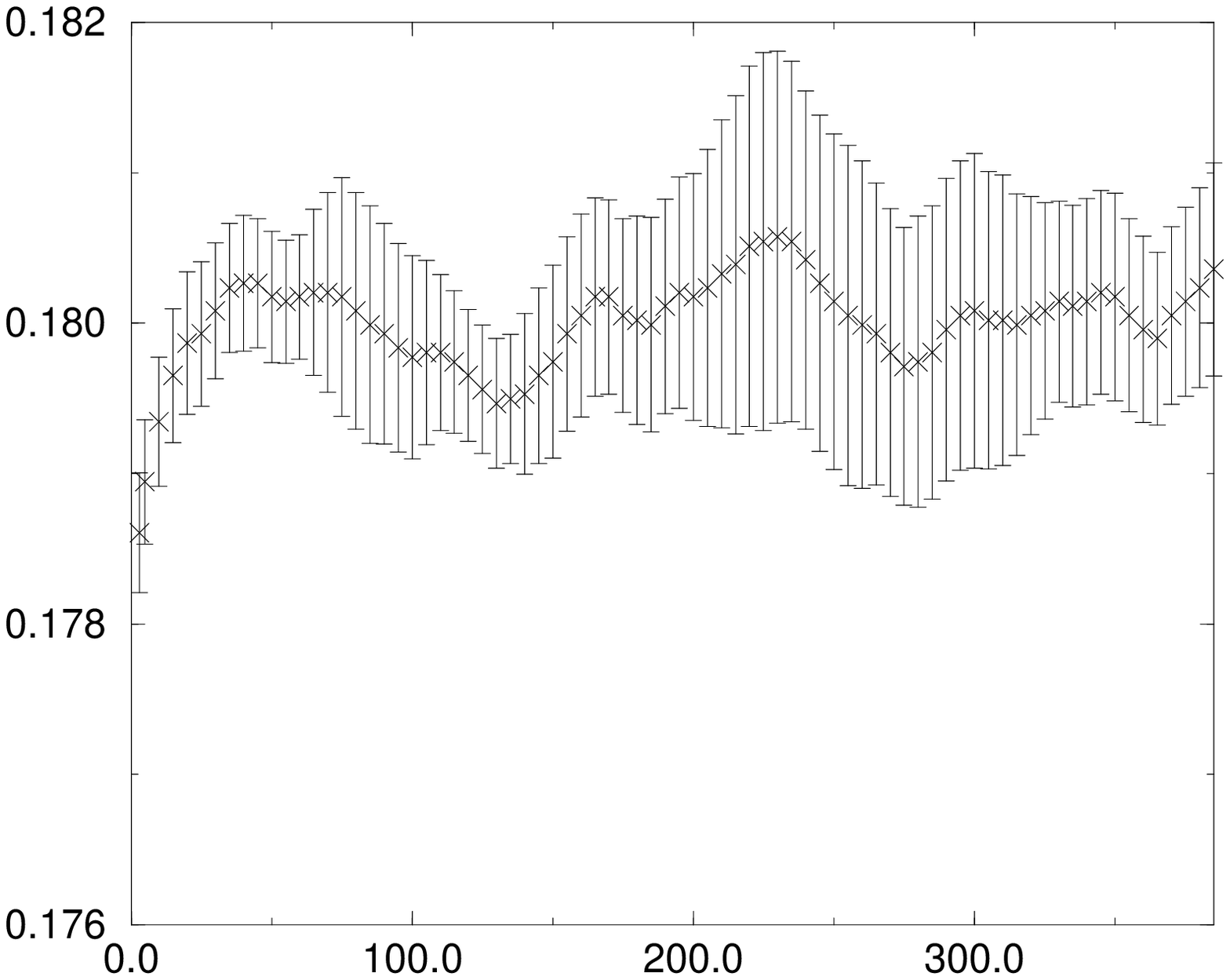}}}
\end{picture}
\caption{  The characteristic function $\varphi(2,m_0)$ estimated with $(L_1,L_2)=(72,144)$
from a 'local' fit of $M_d(t)$ for the Potts model.
For the lattice size $L_2=144$, the initial magnetization $m_0=0.14$.}
\label{f4a}
\end{figure}

\renewcommand{\thefigure}{4b}

\begin{figure}[t]\centering
\epsfysize=12cm
\epsfclipoff
\fboxsep=0pt
\setlength{\unitlength}{1cm}
\begin{picture}(13.6,12)(0,0)
\put(1.2,8.5){\makebox(0,0){$\varphi(2,m_0)$}}
\put(11.8,1.2){\makebox(0,0){$t$}}
\put(6.7,4.5){\makebox(0,0){\footnotesize $m_0=0.30$}}
\put(0,0){{\epsffile{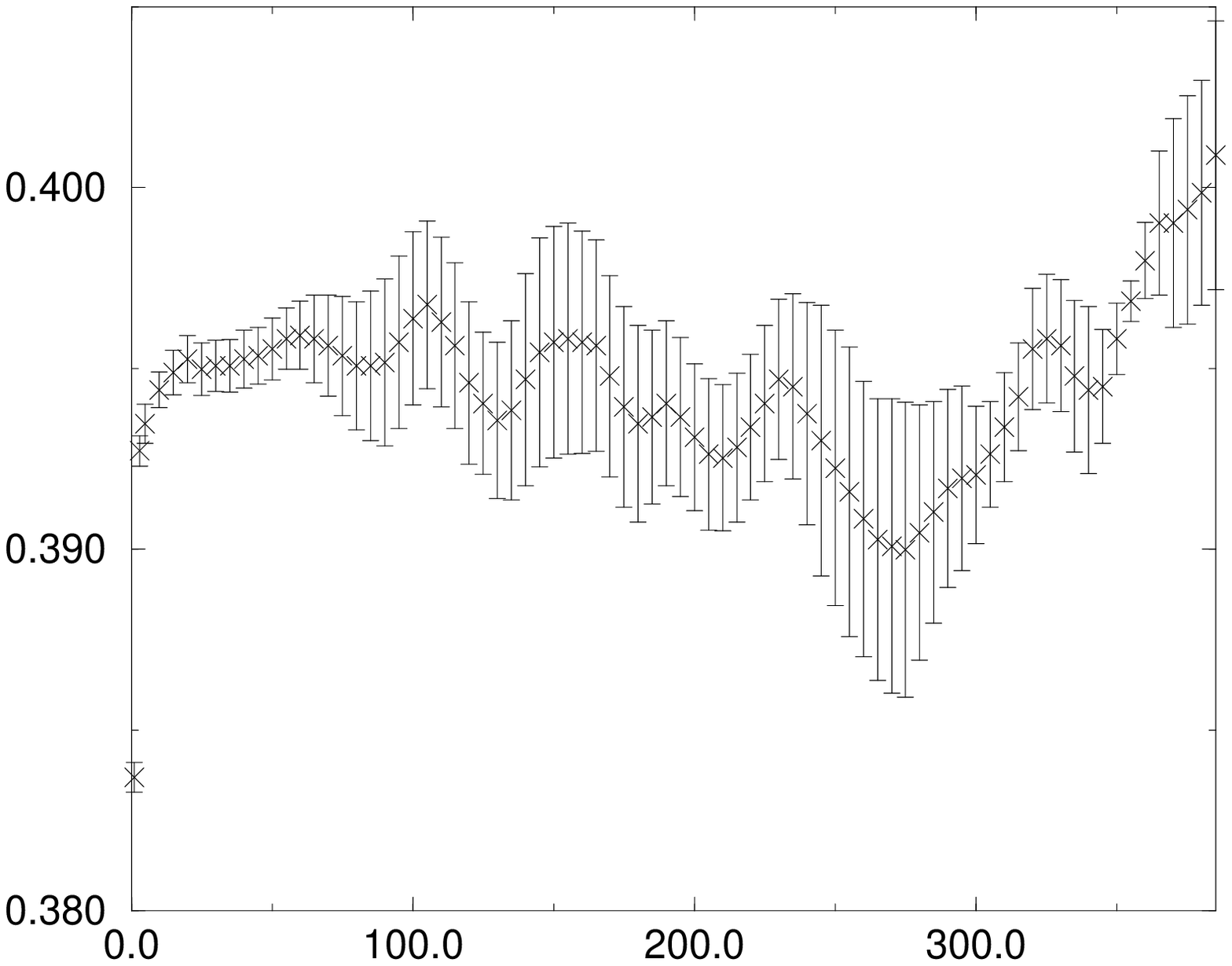}}}
\end{picture}
\caption{ The characteristic function  $\varphi(2,m_0)$ estimated with $(L_1,L_2)=(72,144)$
from a 'local' fit of $M_d(t)$  for the Potts model.
For the lattice size $L_2=144$, the initial magnetization $m_0=0.30$.
}
\label{f4b}
\end{figure}

\renewcommand{\thefigure}{4c}

\begin{figure}[t]\centering
\epsfysize=12cm
\epsfclipoff
\fboxsep=0pt
\setlength{\unitlength}{1cm}
\begin{picture}(13.6,12)(0,0)
\put(1.2,9.0){\makebox(0,0){$\varphi(2,m_0)$}}
\put(11.8,1.2){\makebox(0,0){$t$}}
\put(6.7,5.2){\makebox(0,0){\footnotesize $m_0=0.60$}}
\put(0,0){{\epsffile{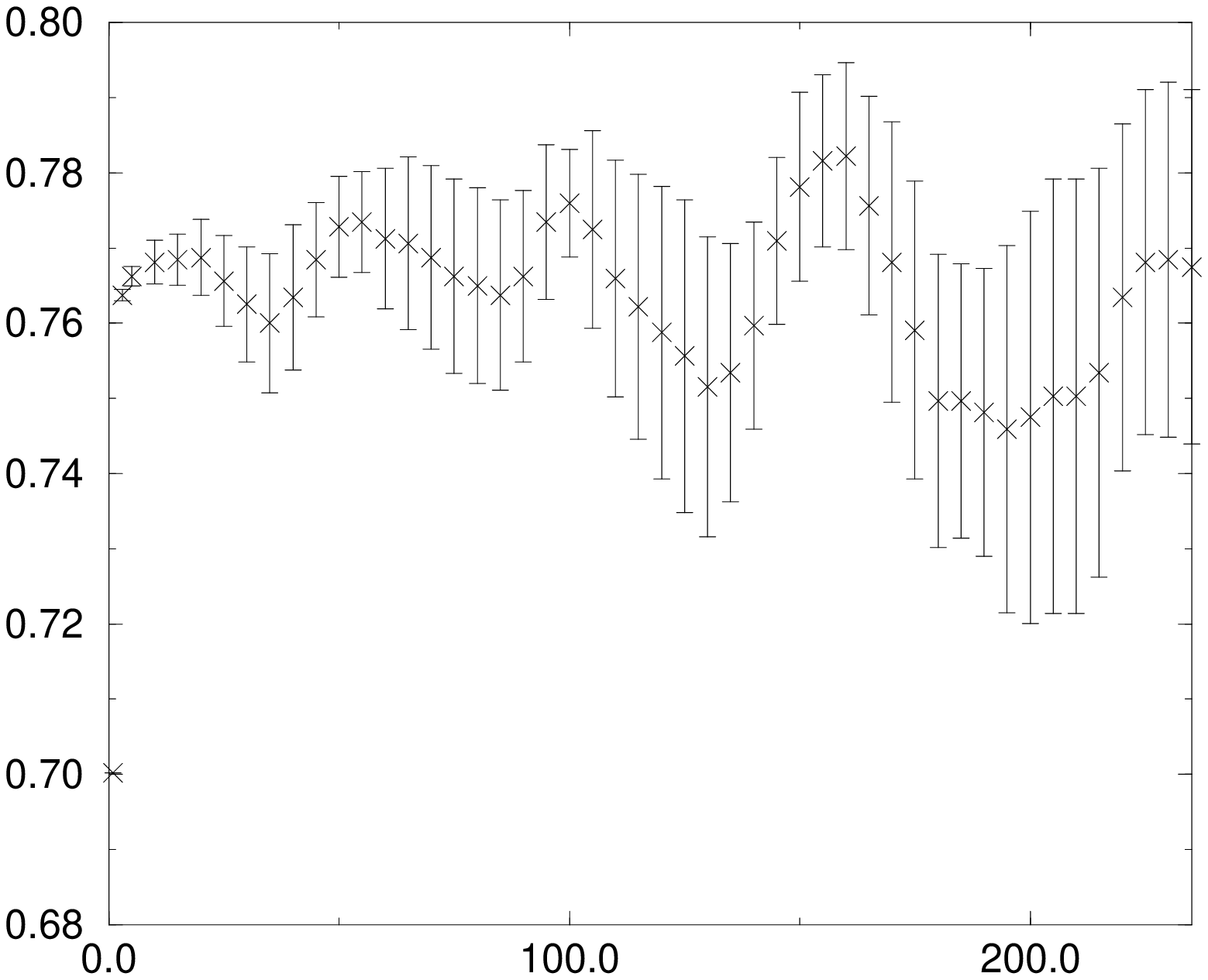}}}
\end{picture}
\caption{  The characteristic function  $\varphi(2,m_0)$ estimated with $(L_1,L_2)=(72,144)$
from a 'local' fit of $M_d(t)$  for the Potts model.
For the lattice size $L_2=144$, the initial magnetization $m_0=0.60$.
}
\label{f4c}
\end{figure}

\renewcommand{\thefigure}{5}

\begin{figure}[t]\centering
\epsfysize=12cm
\epsfclipoff
\fboxsep=0pt
\setlength{\unitlength}{1cm}
\begin{picture}(13.6,12)(0,0)
\put(1.2,8.8){\makebox(0,0){$U_d(t)$}}
\put(11.8,1.2){\makebox(0,0){$t$}}
\put(8.4,5.5){\makebox(0,0){\footnotesize$L=144, m_0=0.14$}}
\put(10.,7.5){\makebox(0,0){\footnotesize$L=72, m_0=0.175$}}
\put(7.2,9.5){\makebox(0,0){\footnotesize$L=72, m_0=0.185$}}
\put(0,0){{\epsffile{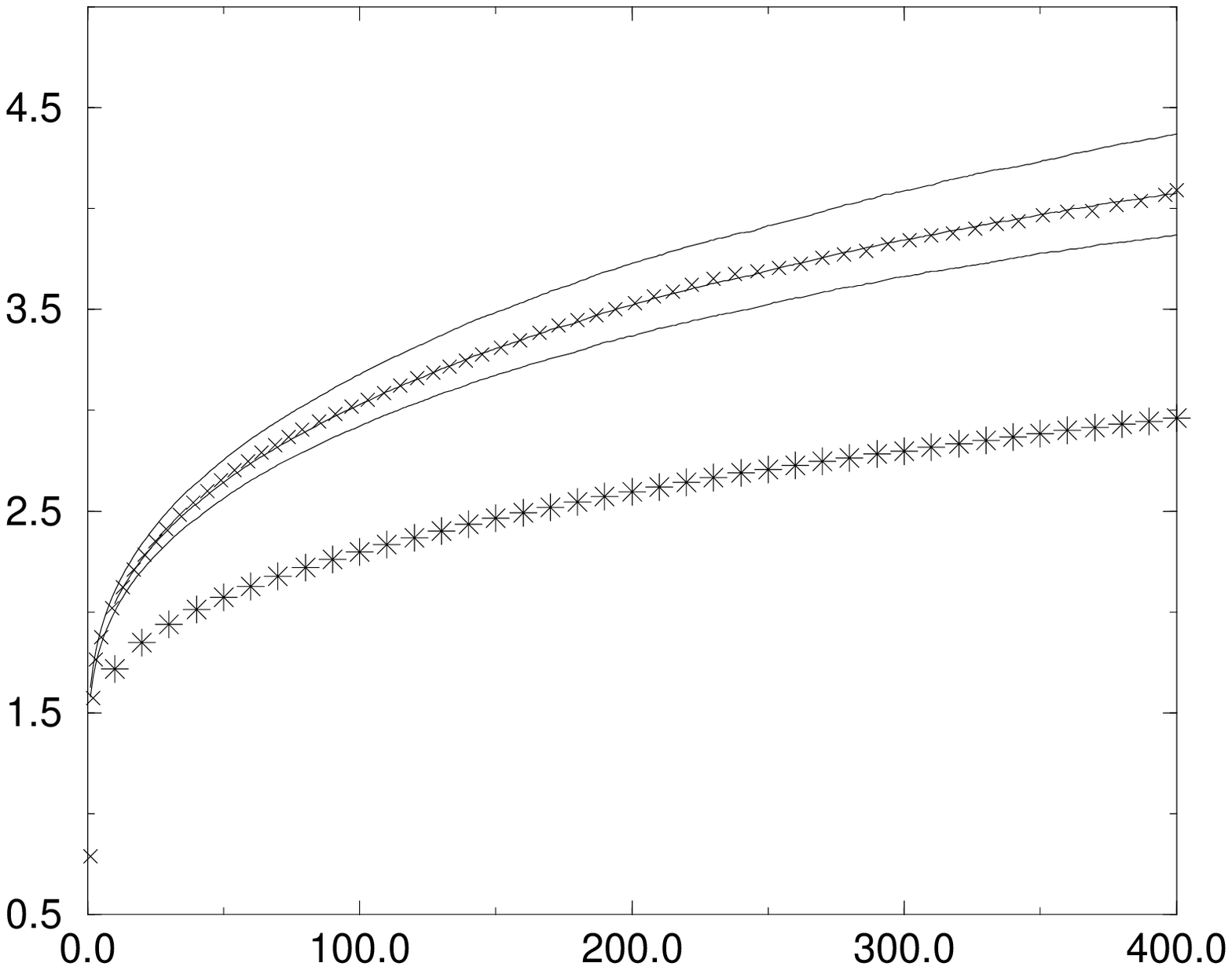}}}
\end{picture}
\caption{ The scaling plot for the Binder-type cumulant
$U_d(t,m_0)$ with $(L_1,L_2)=(72,144)$ for the Potts model.
For the lattice size $L_2=144$, the initial magnetization $m_0=0.14$.}
\label{f5}
\end{figure}

\renewcommand{\thefigure}{6}

\begin{figure}[t]\centering
\epsfysize=12cm
\epsfclipoff
\fboxsep=0pt
\setlength{\unitlength}{1cm}
\begin{picture}(13.6,12)(0,0)
\put(1.2,9.5){\makebox(0,0){$x(b,m_0)$}}
\put(12.2,1.2){\makebox(0,0){$m_0$}}
\put(10.8,8.){\makebox(0,0){\footnotesize$b=2$}}
\put(8.8,5.2){\makebox(0,0){\footnotesize$b=4$}}
\put(0,0){{\epsffile{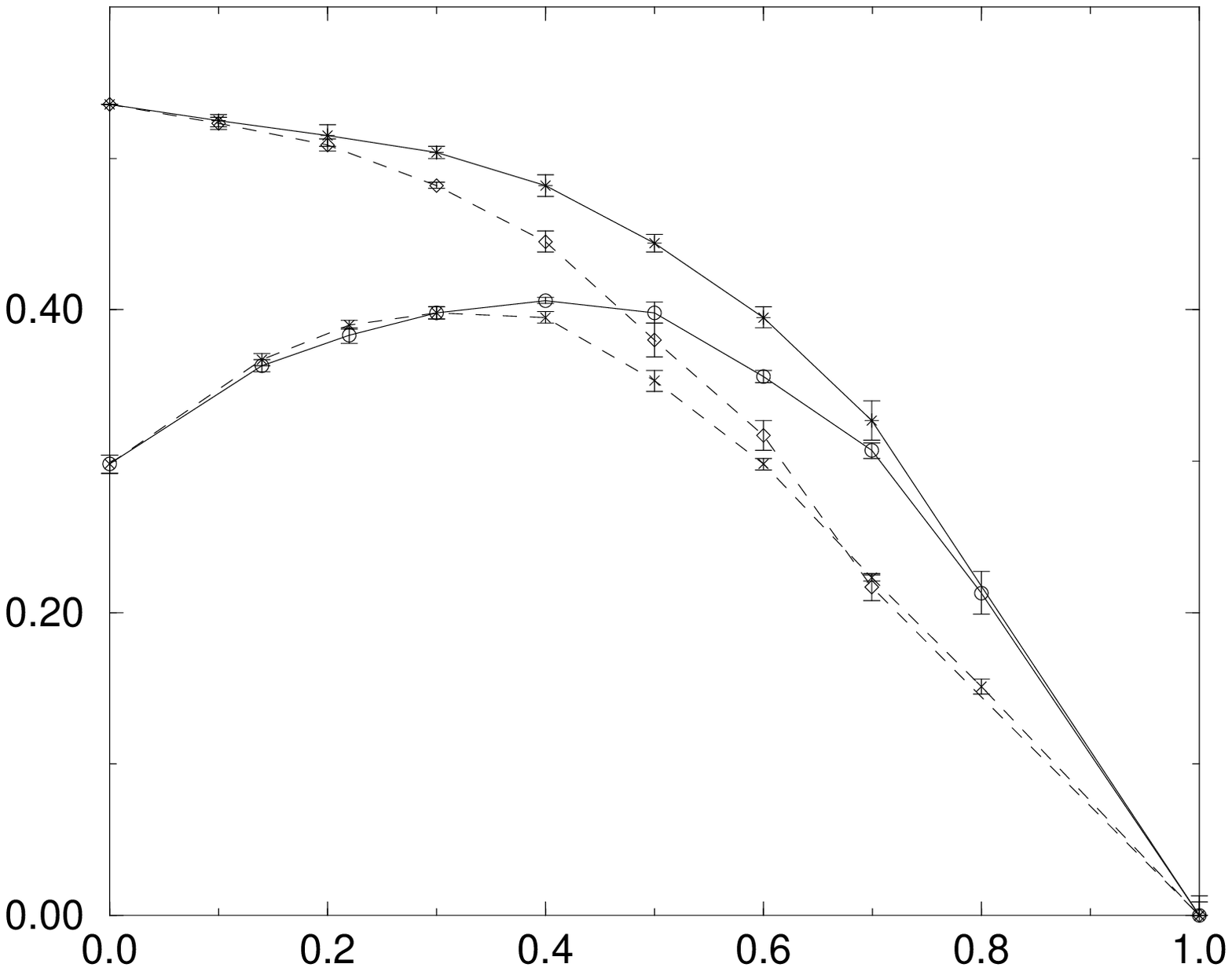}}}
\end{picture}
\caption{ The effective dimension $x(b,m_0)$ for both the Ising
and the Potts model. For the Potts model the data are
obtained from the magnetization difference $M_d(t,m_0)$
with $(L_1,L_2)=(72,144)$ and $(36,144)$. Circles with solid line
 are of $x(2,m_0)$
and crosses with dashed line are of $x(4,m_0)$. 
For the Ising model results are obtained from the magnetization
$M(t,m_0)$ with $(L_1,L_2)=(64,128)$ and $(32,128)$.
Stars with solid line
 are of $x(2,m_0)$ and diamonds with dashed line are of $x(4,m_0)$.
}
\label{f6}
\end{figure}

\renewcommand{\thefigure}{7}

\begin{figure}[t]\centering
\epsfysize=12cm
\epsfclipoff
\fboxsep=0pt
\setlength{\unitlength}{1cm}
\begin{picture}(13.6,12)(0,0)
\put(1.2,8.2){\makebox(0,0){$M(t)$}}
\put(11.8,1.2){\makebox(0,0){$t$}}
\put(8.4,7.){\makebox(0,0){\footnotesize$L=144, m_0=0.14$}}
\put(10.,4.8){\makebox(0,0){\footnotesize$L=72, m_0=0.175$}}
\put(9.2,3.1){\makebox(0,0){\footnotesize$L=72, m_0=0.185$}}
\put(0,0){{\epsffile{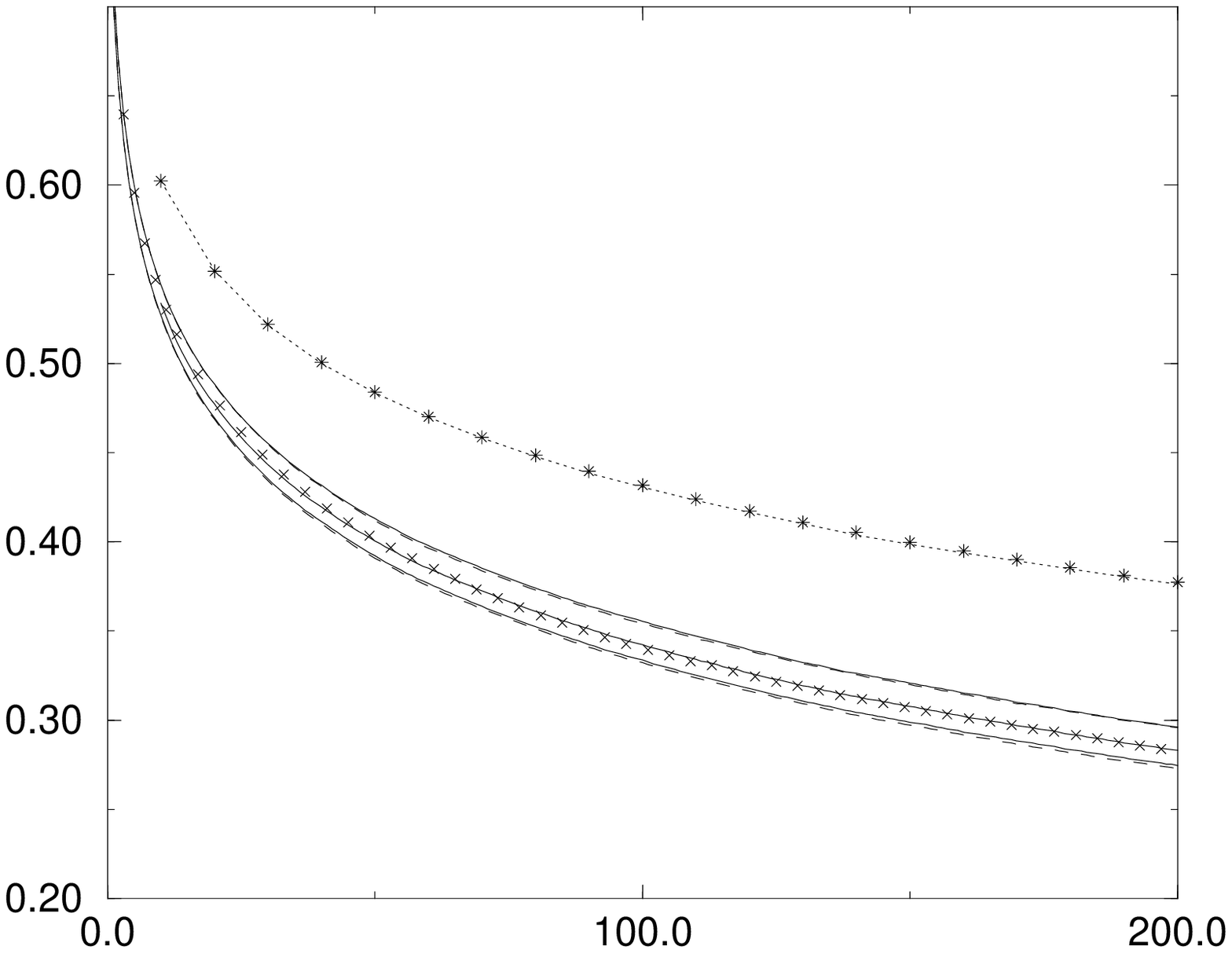}}}
\end{picture}
\caption{ The scaling plot for the magnetization difference
$M_d(t)$ with $(L_1,L_2)=(72,144)$
for the Potts model.
For the lattice size $L_2=144$, the initial magnetization $m_0=0.14$.
The sharp preparation technique is not adopted. 
For comparison, the corresponding magnetization difference profiles 
with the sharp preparation technique are also
displayed for the lattice
sizes $L=144$ and $L=72$ with the dotted line and the dashed lines
respectively.
}
\label{f7}
\end{figure}

\renewcommand{\thefigure}{8}

\begin{figure}[t]\centering
\epsfysize=12cm
\epsfclipoff
\fboxsep=0pt
\setlength{\unitlength}{1cm}
\begin{picture}(13.6,12)(0,0)
\put(1.2,8.0){\makebox(0,0){$M(t)$}}
\put(11.8,1.2){\makebox(0,0){$t$}}
\put(5.,4.1){\makebox(0,0){\footnotesize$L=128, m_0=0.40$}}
\put(4.5,5.5){\makebox(0,0){\footnotesize$L=64, m_0=0.53$}}
\put(8.2,7.){\makebox(0,0){\footnotesize$L=64, m_0=0.59$}}
\put(0,0){{\epsffile{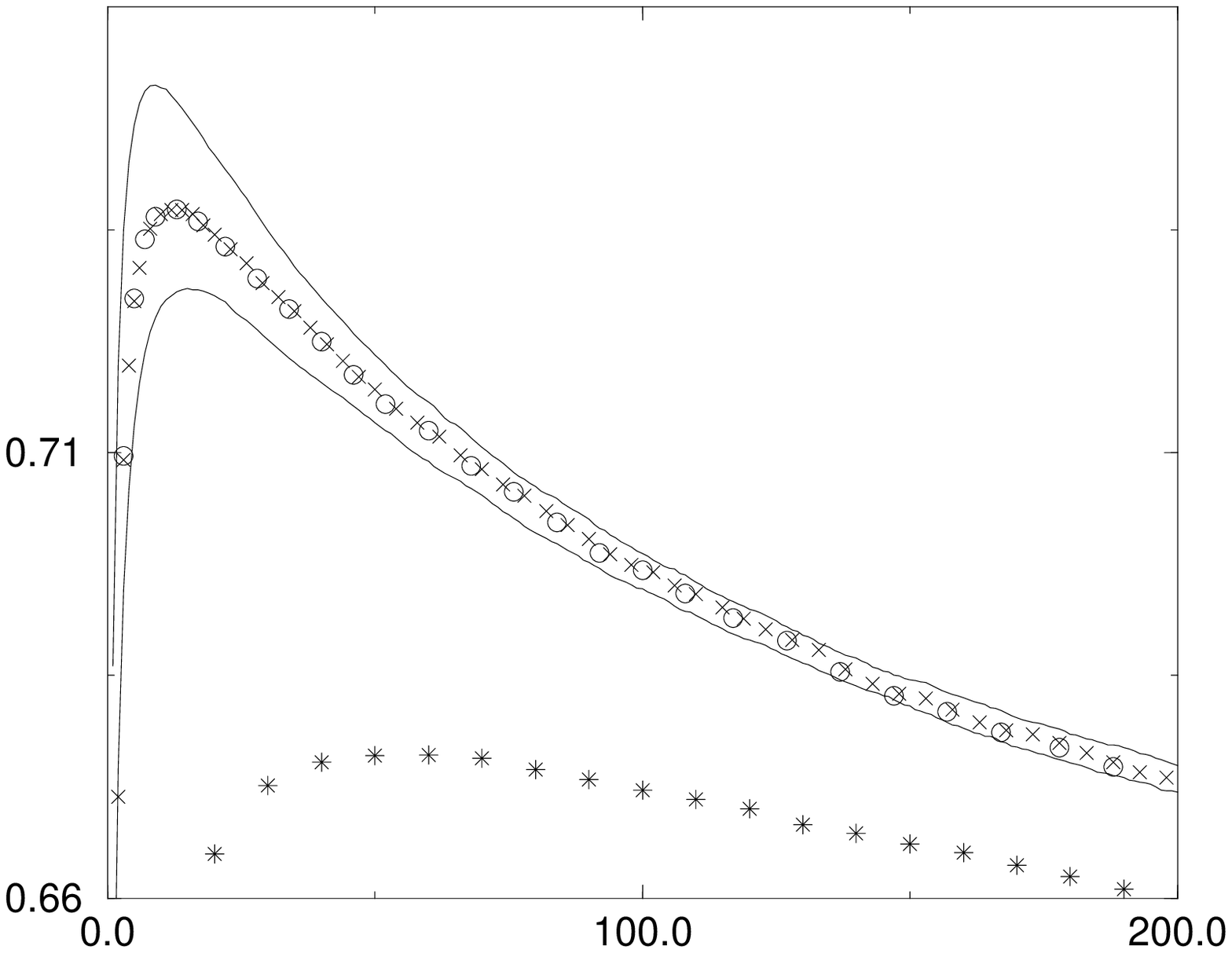}}}
\end{picture}
\caption{ The scaling plot for the magnetization with $(L_1,L_2)=(64,128)$
for the Ising model.
Crosses correspond to $z=2.15$ and circles are from $z=2.17$.}
\label{f8}
\end{figure}

\renewcommand{\thefigure}{9}

\begin{figure}[t]\centering
\epsfysize=12cm
\epsfclipoff
\fboxsep=0pt
\setlength{\unitlength}{1cm}
\begin{picture}(13.6,12)(0,0)
\put(1.2,8.0){\makebox(0,0){$M(t)$}}
\put(11.8,1.2){\makebox(0,0){$t$}}
\put(6.,5.5){\makebox(0,0){\footnotesize$L=32, m_0=0.55$}}
\put(9.7,7.){\makebox(0,0){\footnotesize$L=32, m_0=0.61$}}
\put(0,0){{\epsffile{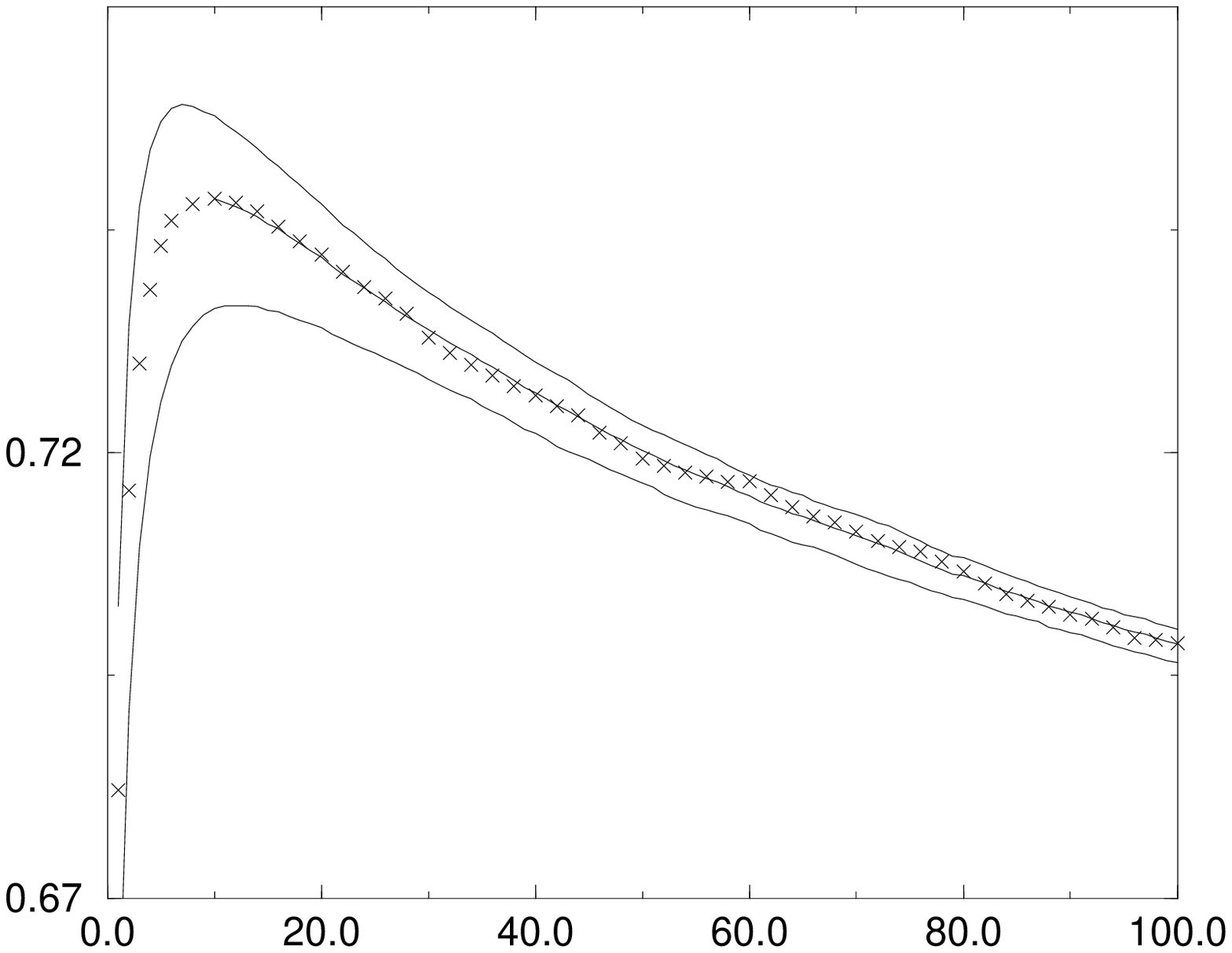}}}
\end{picture}
\caption{ The scaling plot for the magnetization with $(L_1,L_2)=(32,128)$
for the Ising model.
Crosses are for the rescaled magnetization for $L=128$ with $m_0=0.30$.}
\label{f9}
\end{figure}

\renewcommand{\thefigure}{10}

\begin{figure}[t]\centering
\epsfysize=12cm
\epsfclipoff
\fboxsep=0pt
\setlength{\unitlength}{1cm}
\begin{picture}(13.6,12)(0,0)
\put(1.2,8.5){\makebox(0,0){$\varphi(2,m_0)$}}
\put(11.8,1.2){\makebox(0,0){$t$}}
\put(6.7,4.5){\makebox(0,0){\footnotesize$m_0=0.40$}}
\put(0,0){{\epsffile{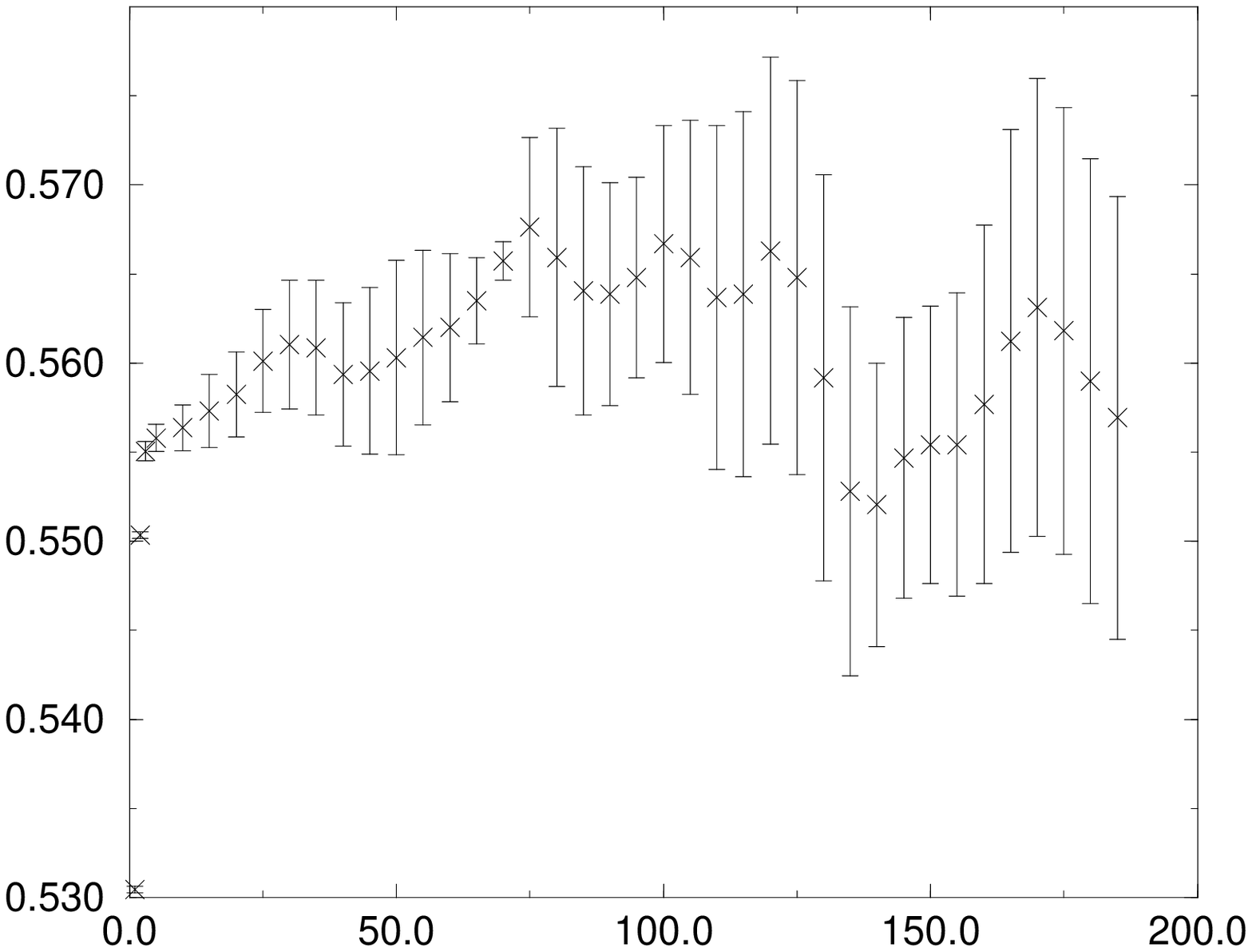}}}
\end{picture}
\caption{ The characteristic function $\varphi(2,m_0)$ estimated with $(L_1,L_2)=(64,128)$
from a 'local' fit of $M(t)$ for the Ising model.
For the lattice size $L_2=128$, the initial magnetization $m_0=0.40$.}
\label{f10}
\end{figure}

\renewcommand{\thefigure}{11}

\begin{figure}[t]\centering
\epsfysize=12cm
\epsfclipoff
\fboxsep=0pt
\setlength{\unitlength}{1cm}
\begin{picture}(13.6,12)(0,0)
\put(1.2,9.5){\makebox(0,0){$y(2,m_0)$}}
\put(12.2,1.2){\makebox(0,0){$m_0$}}
\put(7.,9.){\makebox(0,0){\footnotesize$Ising\ model$}}
\put(9.3,5.2){\makebox(0,0){\footnotesize$Potts\ model$}}
\put(0,0){{\epsffile{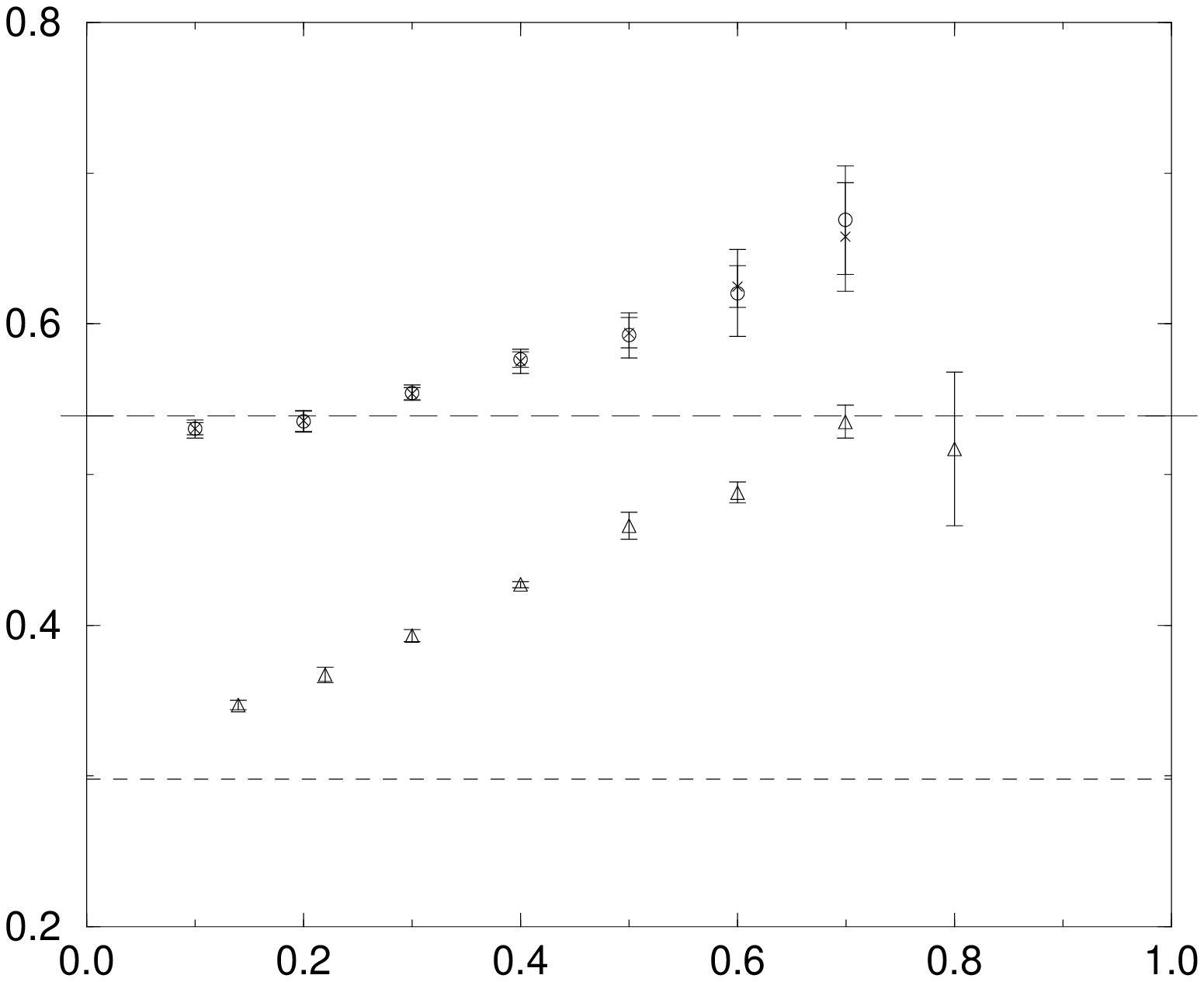}}}
\end{picture}
\caption{ The effective dimension $y(2,m_0)$ 
for both the Ising and the Potts model.
The long dashed and dashed line are the exponent $x_0$
for the Ising and the Potts model respectively,
measured in the limit $m_0=0.0$.
Crosses are of $y(2,m_0)$ for the Ising model
calculated from Table 2
 and circles are the same but obtained from
the scaling plot of $M_d(t,m_0)$.
Triangles are of $y(2,m_0)$ for the Potts model
obtained from Table 1.
}
\label{f11}
\end{figure}

\renewcommand{\thefigure}{12}

\begin{figure}[t]\centering
\epsfysize=12cm
\epsfclipoff
\fboxsep=0pt
\setlength{\unitlength}{1cm}
\begin{picture}(13.6,12)(0,0)
\put(1.2,9.){\makebox(0,0){$y(2,m_0)$}}
\put(12.2,1.2){\makebox(0,0){$m_0$}}
\put(7.,9.){\makebox(0,0){\footnotesize$Ising\ model$}}
\put(0,0){{\epsffile{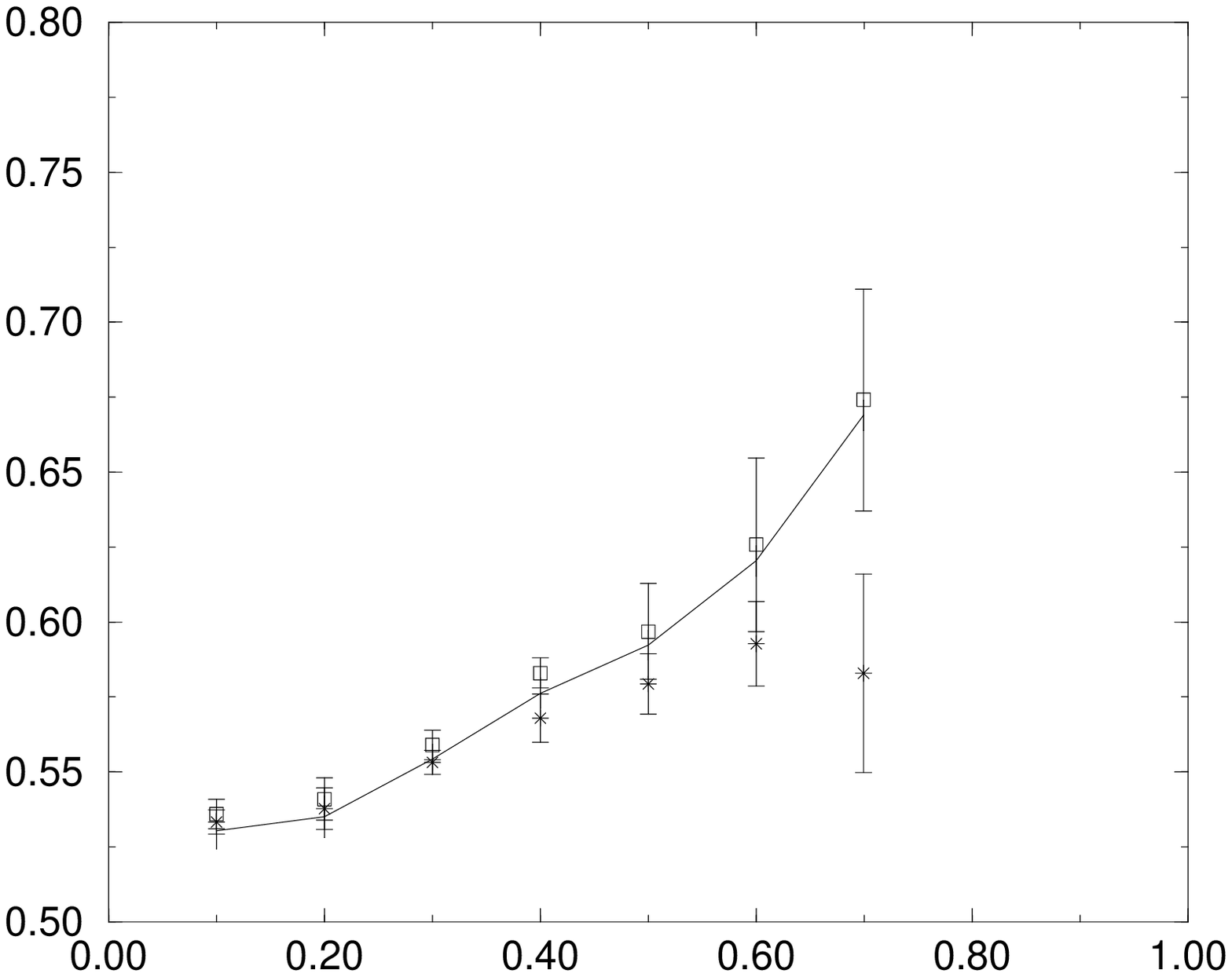}}}
\end{picture}
\caption{ The effective dimension $y(2,m_0)$ 
for the Ising model with $z=2.17$.
The long dashed line is the exponent $x_0$
for the Ising model.
Stars are of $y(2,m_0)$ 
calculated from $M(t,m_0)$
 and the squares are the same but obtained from
the scaling plot of $M_d(t,m_0)$.
The solid line is that estimated from $M_d(t,m_0)$
but with $z=2.15$.
}
\label{f12}
\end{figure}

\end{document}